\documentclass[prb,superscriptaddress,showpacs,floatfix,tightenlines,10pt,twocolumn]{revtex4}
\usepackage{amssymb}
\usepackage{amsmath}
\usepackage{graphicx}

\begin{document}

\title{Orbital and interlayer Skyrmions crystals in bilayer graphene}
\author{R. C\^{o}t\'{e}}
\affiliation{D\'{e}partement de physique, Universit\'{e} de Sherbrooke, Sherbrooke, Qu%
\'{e}bec, J1K 2R1, Canada}
\author{Wenchen Luo }
\affiliation{D\'{e}partement de physique, Universit\'{e} de Sherbrooke, Sherbrooke, Qu%
\'{e}bec, J1K 2R1, Canada}
\author{Branko Petrov}
\affiliation{D\'{e}partement de physique, Universit\'{e} de Sherbrooke, Sherbrooke, Qu%
\'{e}bec, J1K 2R1, Canada}
\author{Yafis Barlas}
\affiliation{National High Magnetic Field Laboratory and Department of Physics, The
Florida State University, Tallahassee, Florida 32306, USA}
\author{A. H. MacDonald}
\affiliation{Department of Physics, The University of Texas at Austin, Austin, Texas
78712, USA}

\begin{abstract}
A graphene bilayer in a transverse magnetic field has a set of Landau levels
with energies $E=\pm \sqrt{N\left( N+1\right) }\hslash \omega _{c}^{\ast }$
where $\omega _{c}^{\ast }$ is the effective cyclotron frequency and $%
N=0,1,2,...$ All Landau levels but $N=0$ are four times degenerate counting
spin and valley degrees of freedom. The Landau level $N=0$ has an extra
degeneracy due to the fact that orbitals $n=0$ and $n=1$ both have zero
kinetic energies. At integer filling factors, Coulomb interactions produce a
set of broken-symmetry states with partial or full alignement in space of
the valley and orbital pseudospins. These quantum Hall pseudo-ferromagnetic
states support topological charged excitations in the form of orbital and
valley Skyrmions. Away from integer fillings, these topological excitations
can condense to form a rich variety of Skyrme crystals with interesting
properties. We study in this paper different crystal phases that occur when
an electric field is applied between the layers. We show that orbital
Skyrmions, in analogy with spin Skyrmions, have a texture of electrical
dipoles that can be controlled by an in-plane electric field. Moreover, the
modulation of electronic density in the crystalline phases are
experimentally accessible through a measurement of their local density of
states
\end{abstract}

\date{\today }
\maketitle

\section{INTRODUCTION}

Bilayer graphene is a system consisting of two layers of graphene separated
by a distance $d=3.337$ \AA . In the Bernal stacking structure, one of the
two honeycomb sublattice sites in each layer has a near neighbor in the
other layer and one does not. In a transverse magnetic field, the
two-dimensional electron gas (2DEG) develops a set of Landau levels with
energies $E^{0}=\pm \sqrt{N\left( N+1\right) }\hslash \omega _{c}^{\ast }$
where $N=0,1,2,...$ and $\omega _{c}^{\ast }=eB/m^{\ast }c$ is the effective
cyclotron frequency. The effective mass given by $m^{\ast }=2\hslash
^{2}\gamma _{1}/3\gamma _{0}^{2}a_0^{2}$ where $a_0$ is the lattice constant of
graphene and $\gamma _{0}$ and $\gamma _{1}$ are in-plane nearest-neighbor
and inter-plane hopping parameters. By comparison, the effective mass is
zero in graphene and the Landau levels energies are then given by $E^{0}=\pm 
\sqrt{2}\hslash v_{F}\sqrt{N}/\ell $ where $\ell ^{2}=\hslash c/eB$ is the
magnetic length and $v_{F}=\sqrt{3}a_0\gamma _{0}/2\hslash $ is the Fermi
velocity.

In the absence of an electric field between the layers and Zeeman coupling,
the Landau level $N=0$ in a graphene bilayer contains 8 states for each
guiding center orbital. The extra degeneracy is due to the fact that \textit{%
orbitals} (we define this term in the next section) $n=0$ and $n=1$ both
have the same kinetic energy $E=0.$ Consequently, an electron in $N=0$ must
be described by its spin, valley (or layer), and orbital quantum numbers in
addition to its guiding center index $X$ in the Landau gauge. When Coulomb
interaction is considered, this extra degeneracy produces a rich phase
diagram for the bilayer graphene's 2DEG. More so, in fact, than in a
semiconductor 2DEG. In a series of related papers\cite{yafis1,yafis2,cote1},
we have shown that the octet degeneracy is lifted by the Coulomb
interaction. The broken-symmetry ground states that emerge can be described
as quantum Hall pseudo-ferromagnets in the pseudospin language where
fictitious spins are associated with the valley and orbital indices. These
new states have interesting transport properties such as an
intra-Landau-level cyclotron mode and a layer pseudospin with a quadratic ($%
\omega \sim q^{2}$) dispersion implying a vanishing superfluid density.

It is well known that a quantum Hall ferromagnet (QHF) in a usual
semiconductor 2DEG has topological excitations named spin Skyrmions\cite%
{skyrmionreview}. A single Skyrmion spin texture has its spins aligned with
the Zeeman field at infinity, reversed at the center of the Skyrmion, and
has non zero $XY$ spin components at intermediate distance which have a
vortex-like configuration. Skyrmions carry electric charge. Calculations
have shown that Skyrmion-anti-Skyrmion pairs have lower energy than
electron-hole quasiparticles near filling factor $\nu =1$ and dominate the
transport properties of the QHF\cite{fertigskyrmion}. A quantum Hall bilayer
has, in addition to spin Skyrmions, topological excitations named pseudospin
Skyrmions where the fictitious spin is associated with the layer index. We
show in this paper that, in a graphene bilayer, there is a third
possibility: that of an orbital-pseudospin Skyrmion. This quasiparticle has
an associated electric dipole texture in the plane of the layers and can be
seen as the analog of a spin Skyrmion who carries a magnetic texture.

We present several crystal states with pseudospin texture that occur near
integer filling factors in Landau level $N=0.$ We assume full spin
polarization of the 2DEG and concentrate on Skyrme crystals with valley
and/or orbital pseudospin textures. We allow for the presence of an electric
field between the layers that creates a charge imbalance and we study the
evolution of the Skyrme crystals as a function of this electrical ``bias'' for
different filling factors. Our goal is not to study the full phase diagram
of the bilayer but to focus on a small number of interesting crystal phases
that are likely to occur near integer filling factors. Our calculation shows
that Skyrmions tend to crystallize in pairs and that an orbital-pseudospin
texture is favored over valley-pseudospin texture at filling factors $\nu
=-3,-1,1,3.$ Valley-pseudospin Skyrmion crystals occur at filling factor $%
\nu =-2,2$ and involve a texture in both the $n=0$ and $n=1$ orbitals. This
possibility was discussed before, for an isolated Skyrmion, in Ref.~\onlinecite{abanin}. 
We also show that Skyrmions with different electric
charge $q$ can be distinguished on the basis of their density of states.
Moreover, the real-space density pattern of a Skyrmion crystal is
accessible by a measurement of its local density of states.

This paper is organized as follows. In Sec.~II, we present the effective
two-band tight-binding model that we use to describe the graphene bilayer.
In Sec.~III, we derive the Hamiltonian of the bilayer graphene's 2DEG in the
Hartree-Fock approximation. Sec.~IV presents the pseudospin language used to
describe the various crystal phases. In Sec.~V, we introduce spin and
orbital-pseudospin Skyrmions. We then present and discuss various Skyrmion
crystal phases at filling factors $\nu =-3,-2,-1$ (or equivalently $\nu
=1,2,3$) in Sec.~VI. The total and local densities of states are defined in
Sec.~VII and calculated for some of the crystal phases. The electric dipole
texture associated with an orbital Skyrmion crystal is computed in Sec.~ VIII. 
We conclude in Sec.~IX with a discussion of some of the terms
neglected in our simple tight-binding model.

\section{EFFECTIVE HAMILTONIAN\qquad}

\begin{figure}[tbph]
\includegraphics[scale=1]{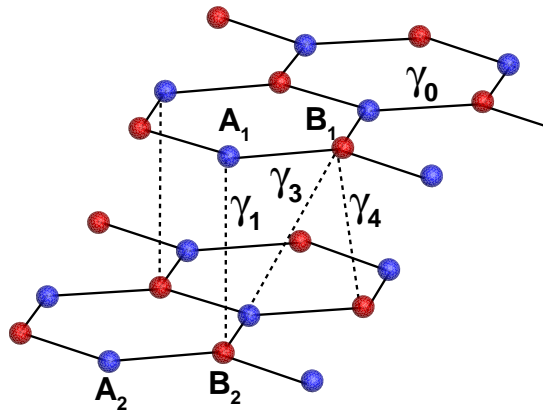}
\caption{(Color online) Crystal structure and definition of the hopping
parameters for the graphene bilayer.}
\label{fig1}
\end{figure}

We consider the graphene bilayer in the Bernal stacking arrangement\cite%
{castronetoreview} represented in Fig.~\ref{fig1}. We denote the two basis
atoms of the top layer by $A_{1}$ and $B_{1}$ and those of the bottom layer
by $A_{2}$ and $B_{2}$ with atoms $A_{1}$ situated directly above atoms $%
B_{2}$. The bilayer is placed in an external transverse electric field in
order to control the electrical potential difference (i.e. the ``bias'') $%
\Delta _{B}$ between the layers that causes the charge imbalance. To
simplify our analysis, we assume complete spin polarization of the electron
gas and neglect trigonal warping (the $\gamma _{3}$ hopping in Fig.~\ref%
{fig1}). We also use an effective two-band model\cite{mccann} to describe
the low-energy excitations of the bilayer in a quantizing magnetic field in
the valleys $\mathbf{K}=\left( -4\pi /3a_{0},0\right) $ and $\mathbf{K}%
^{\prime }=\left( 4\pi /3a_{0},0\right) $. Although we will not consider
these terms in the bulk of this paper, we could generalize this model by
including the $\gamma _{4}$ hopping term as well as an additional term $%
\Delta $ representing the difference in the crystal field experienced by the
inequivalent atoms $A$ and $B$ in the same plane. With these approximations,
we get the Hamiltonian:%
\begin{equation}
H_{K}^{0}=\left( 
\begin{array}{cc}
-\frac{\Delta _{B}}{2}+\left( \beta _{0}\Delta _{B}+\zeta _{1}\right)
aa^{\dag } & \zeta _{2}a^{2} \\ 
\zeta _{2}\left( a^{\dag }\right) ^{2} & \frac{\Delta _{B}}{2}+\left( -\beta
_{0}\Delta _{B}+\zeta _{1}\right) a^{\dag }a%
\end{array}%
\right) ,  \label{s_3}
\end{equation}

in the basis $\left( A_{2},B_{1}\right) . $ In Eq.~(\ref{s_3}), $a,a^{\dag }$
are the ladder operators for the Landau levels and we have defined the
parameters%
\begin{eqnarray}
\zeta _{1} &=&2\mathrm{sgn}\left( \gamma _{0}\gamma _{4}\right) \sqrt{\beta
_{0}\beta _{4}}\gamma _{1}+\left( \beta _{0}+\beta _{4}\right) \Delta , \\
\zeta _{2} &=&2\mathrm{sgn}\left( \gamma _{0}\gamma _{4}\right) \sqrt{\beta
_{0}\beta _{4}}\Delta +\left( \beta _{0}+\beta _{4}\right) \gamma _{1},
\end{eqnarray}%
where 
\begin{eqnarray}
\beta _{0} &=&\frac{\hslash \omega _{c}^{\ast }}{\gamma _{1}}, \\
\beta _{4} &=&\left( \frac{\gamma _{4}}{\gamma _{0}}\right) ^{2}\frac{%
\hslash \omega _{c}^{\ast }}{\gamma _{1}},
\end{eqnarray}%
are unitless constants and sgn denotes the signum function. The effective
cyclotron frequency is defined by $\omega _{c}^{\ast }=eB/m^{\ast }c$ with
the effective electronic mass given by $m^{\ast }=2\hslash ^{2}\gamma
_{1}/3\gamma _{0}^{2}a_0^{2}=0.054m_{0}$ where $m_{0}$ is the bare electronic
mass and $a_0=2.46$ \AA\ is the lattice parameter of graphene. Note that, in
the basis $\left( A_{2},B_{1}\right) $, $H_{K^{\prime }}^{0}=\left(
H_{K}^{0}\right) ^{\dag }.$

In the case where $\gamma _{4}=\Delta =\Delta _{B}=0,$ the Landau level
energies are given by%
\begin{equation}
E^{0}=\pm \sqrt{N\left( N+1\right) }\hslash \omega _{c}^{\ast },
\end{equation}%
with $N=0,1,2,3,...$ All Landau levels are four time degenerate (including
spin and valley degrees of freedom) with the exception of $N=0$ that is
eight times degenerate. With finite $\gamma _{4}$,$\Delta $,$\Delta _{B}$,
we find for the spin up states of $N=0$ the following spinors and energies:%
\begin{eqnarray}
\left( 
\begin{array}{c}
0 \\ 
h_{0,X}\left( \mathbf{r}\right)%
\end{array}%
\right) ,\;E_{K,0,X}^{0} &=&\frac{1}{2}\Delta _{B},  \label{s_10} \\
\left( 
\begin{array}{c}
0 \\ 
h_{1,X}\left( \mathbf{r}\right)%
\end{array}%
\right) ,\;E_{K,1,X}^{0} &=&\frac{1}{2}\Delta _{B}-\beta _{0}\Delta
_{B}+\zeta _{1},  \label{s_11}
\end{eqnarray}%
for the $K$ valley and 
\begin{eqnarray}
\left( 
\begin{array}{c}
h_{0,X}\left( \mathbf{r}\right) \\ 
0%
\end{array}%
\right) ,\;E_{K^{\prime},0,X}^{0} &=&-\frac{1}{2}\Delta _{B},  \label{s_12} \\
\left( 
\begin{array}{c}
h_{1,X}\left( \mathbf{r}\right) \\ 
0%
\end{array}%
\right) ,\;E_{K^{\prime},1,X}^{0} &=&-\frac{1}{2}\Delta _{B}+\beta _{0}\Delta
_{B}+\zeta _{1},  \label{s_13}
\end{eqnarray}%
for the $K^{\prime }\ $valley. Note that we have neglected the Zeeman
coupling since we assume complete spin polarization and thus discard the
spin degree of freedom in the rest of our analysis. The functions $%
h_{n,X}\left( \mathbf{r}\right) =e^{-iXy/\ell ^{2}}\varphi _{n}\left(
x-X\right) /\sqrt{L_{y}}$ are the eigenstates in the Landau gauge $\mathbf{A}%
=\left( 0,Bx,0\right) $ with guiding center $X$ and $\varphi _{n}\left(
x\right) $ is the wave function of the one-dimensional harmonic oscillator.
The magnetic length is given by $\ell =\sqrt{\hslash c/eB}=256/\sqrt{B}$ \AA %
. We see from Eqs.~(\ref{s_10}-\ref{s_11}) that, in addition to the spin and
valley quantum numbers, there is in $N=0$ an extra degeneracy due to the
fact that wave functions $h_{0,X}\left( \mathbf{r}\right) $ and $%
h_{1,X}\left( \mathbf{r}\right) $ both have zero kinetic energy if $\Delta
_{B}=0$. Throughout this paper, we will refer to these states as \textit{%
orbitals} $n=0,1$ and use the symbol $N$ for the Landau level index.

\begin{figure}[tbph]
\includegraphics[scale=1]{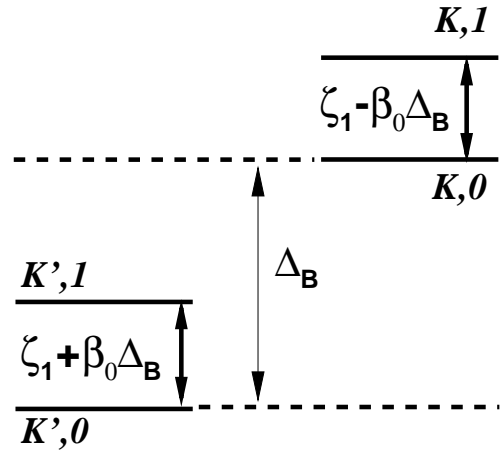}
\caption{Non-interacting energy levels with spin up in Landau level $N=0.$
Note that when $\protect\zeta _{1}=0$, level $K,1$ is below level $K,0$ in
energy.}
\label{fig2}
\end{figure}

From Eqs.~(\ref{s_10})-(\ref{s_13}), we see that, in the Landau level $N=0$,
the electrons are localized on the atoms $A_{2}$ (bottom layer) in the $%
K^{\prime }$ valley and on the atoms $B_{1}$ (top layer) in the $K$ valley.
In $N=0$, the layer index is equivalent to the valley index. An external
electric field lifts both the valley and the orbital degeneracies. The
orbital degeneracy is lifted by the small corrections $\beta _{0}\Delta _{B}$
and $\zeta _{1}$ as shown in Fig.~\ref{fig2}. The values of the intra and
interlayer hoppings are given by $\gamma _{0}=3.12$ eV and $\gamma _{1}=0.39$
eV. The other hopping terms as well as $\Delta $ are not so well known. It
is difficult to get the relative signs of these terms from the litterature.
Recent measurements of these parameters for bilayer graphene give $\gamma
_{4}=0.04-0.07$ and $\Delta =0.005-0.008$ in units of the in-plane hopping $%
\gamma _{0}$. These values are discussed and referenced in Ref.~\onlinecite{castro}. 
Taking the minimal values for $\gamma _{4}$ and $\Delta 
$, we have $\beta _{0}=8.\,\allowbreak 86\times 10^{-3}B,\beta
_{4}=1.\,\allowbreak 31\times 10^{-5}B$ (with the magnetic field in Tesla)
and $\Delta =0.0156$ eV so that $\zeta _{1}=4.\,\allowbreak 042\times
10^{-4}B$ eV.

We showed in Refs.~\onlinecite{yafis1,yafis2,cote1} that when $\zeta _{1}=0$%
, the phase diagram of the 2DEG at integer filling factors $\nu \in \left[
-3,4\right] $ contains phases with interlayer and/or inter-orbital
coherences. Because of the small interlayer spacing ($d=3.337$ \AA ) in a
graphene bilayer, interlayer coherence is rapidly lost when $\Delta _{B}$
increases i.e. for $\Delta _{B}/\left( e^{2}/\kappa \ell \right) \gtrsim
0.001$ according to our numerical calculations ($\kappa $ is the effective
dielectric constant at the position of the graphene layers). Above this
value, inter-orbital coherence sets in when $E_{K,0,X}^{0}>E_{K,1,X}^{0}$%
. From Eqs.~(\ref{s_12}-\ref{s_13}), this is only possible at $\nu =-1,3$.
Indeed, our calculations show that the phase diagram for $\nu =-3,1$ has no
orbital-coherent phase (when $\zeta_{1}=0$) if the band parameters we use are correct.

The precise values of the bias for the transitions between the different
liquid phases at integer filling factors are very sensitive to the exact
values of the hopping parameters. The same is true for the boundaries
between various crystal phases at non-integer filling factors. Moreover, the
number of possible crystal phases is much larger than the number of possible
liquid phases when one considers the various crystal lattices and the
possibility of having more than one electron per unit cell with interlayer
and/or orbital pseudospin textures. For this reason, we focus, in this
paper, on the analysis of a few crystal phases with orbital or interlayer
texture which are likely to appear in the phase diagram of the 2DEG in some
range of values of $\zeta _{1}.$ We assume $\zeta _{1}=0$ for all our
calculations and discuss in the conclusion how the phase diagram is likely
to be changed when $\zeta _{1}\neq 0.$ In our opinion reliable determination
of the phase boundary characterizing the many possible crystalline phases
will require experimental input.

Note that in the absence of Landau level mixing and when maximal spin
polarization is assumed, the ground states at filling factors $\nu =-3,-2,-1$
are equivalent to those at filling factors $\nu =1,2,3.$ It is thus
sufficient for us to study the first three states $\nu =-3,-2,-1.$ If the
approximation of maximal spin polarization is not made, phases with reduced
polarization become possible as the bias is increased. (The energy of half
the spin down (up) states decreases (increases) with bias and levels
crossing do occur). The ground states at $\nu =-3,-2,-1$ are no longer
equivalent to those at $\nu =1,2,3$. The phase diagram is much more complex
in this case but our calculations show that phases with orbital or
interlayer coherences are still present.

\section{HARTREE-FOCK DESCRIPTION OF THE CRYSTAL PHASES}

We now add the Coulomb interaction to the non-interacting Hamiltonian.
We assume that the magnetic field is strong enough so that we
can neglect Landau level mixing. The Hartree-Fock Hamiltonian for the 2DEG
in Landau level $N=0$ is then given by

\begin{gather}
H_{HF}=N_{\varphi } \sum_{n}\sum_{a}E_{a,n}\rho _{n,n}^{a,a}\left( 0\right)
\label{HFhamiltonian} \\
+N_{\varphi }\sum_{a,b}\sum_{n_{1},...,n_{4}}\overline{\sum_{\mathbf{q}}}%
H_{n_{1},n_{2},n_{3},n_{4}}^{a,b}\left( \mathbf{q}\right)  \notag \\
\times \left\langle \rho _{n_{1},n_{2}}^{a,a}\left( -\mathbf{q}\right)
\right\rangle \rho _{n_{3},n_{4}}^{b,b}\left( \mathbf{q}\right)  \notag \\
-N_{\varphi }\sum_{a,b}\sum_{n_{1},...,n_{4}}\sum_{\mathbf{q}%
}X_{n_{1},n_{4},n_{3},n_{2}}^{a,b}\left( \mathbf{q}\right)  \notag \\
\times \left\langle \rho _{n_{1},n_{2}}^{a,b}\left( -\mathbf{q}\right)
\right\rangle \rho _{n_{3},n_{4}}^{b,a}\left( \mathbf{q}\right) ,  \notag
\end{gather}%
where $N_{\varphi }=S/2\pi \ell ^{2}$ is the Landau level degeneracy ($S$ is
the 2DEG area) and all energies are now measured in units of $e^{2}/\kappa
\ell $. The single-particle energies $E_{a,n}$ include capacitive
contributions and are defined by 
\begin{equation}
E_{a,n}=\frac{1}{2}a\Delta _{B}-a\beta _{0}\Delta _{B}n+\left[ \frac{%
\widetilde{\nu }}{2}\frac{d}{\ell }-\widetilde{\nu }_{a}\frac{d}{\ell }%
\right] +\zeta _{1},  \label{ener}
\end{equation}%
with $a,b=\pm 1$ the valley (or equivalently layer) index and $n=0,1$ the
orbital index. (Our convention is that $a=1(-1)$ for the $K(K')$ valley).
Because we work in Landau level $N=0$ only, we define $%
\widetilde{\nu }=\nu +4\in \left[ 0,8\right] $ as the number of filled
levels in $N=0$. In deriving Eq.~(\ref{HFhamiltonian}), we have taken into
account a neutralizing positive background and put the capacitive energy in
the third term on the right-hand side of Eq.~(\ref{ener}). It follows that
the $\mathbf{q}=0$ contribution is absent in the Hartree term of Eq.~(\ref%
{HFhamiltonian}). This convention is indicated by the bar over the summation.

In Eq.~(\ref{HFhamiltonian}), the density operator is defined by

\begin{eqnarray}
\rho _{n_{1},n_{2}}^{a,b}\left( \mathbf{q}\right) &=&\frac{1}{N_{\varphi }}%
\sum_{X_{1},X_{2}}e^{-\frac{i}{2}q_{x}\left( X_{1}+X_{2}\right) } \\
&&\times c_{a,X_{1},n_{1}}^{\dagger }c_{b,X_{2},n_{2}}\delta
_{X_{1},X_{2}+q_{y}\ell ^{2}},  \notag
\end{eqnarray}%
where $c_{a,X,n}^{\dagger }$ creates an electron in the state $\left(
a,X,n\right) $ in the Landau gauge. The intralayer $\left(
H,X=H^{a,a},X^{a,a}\right) $ and interlayer $\left( \widetilde{H},\widetilde{%
X}=H^{a\neq b},X^{a\neq b}\right) $ Hartree and Fock interactions are
defined by 
\begin{equation}
H_{n_{1},n_{2},n_{3},n_{4}}\left( \mathbf{q}\right) =\frac{1}{q\ell }%
K_{n_{1},n_{2}}\left( \mathbf{q}\right) K_{n_{3},n_{4}}\left( -\mathbf{q}%
\right) ,  \label{hart1p}
\end{equation}%
\begin{equation}
X_{n_{1},n_{2},n_{3},n_{4}}\left( \mathbf{q}\right) =\int \frac{d\mathbf{p}%
\ell ^{2}}{2\pi }H_{n_{1},n_{2},n_{3},n_{4}}\left( \mathbf{p}\right) e^{i%
\mathbf{q}\times \mathbf{p}\ell ^{2}},  \label{fock1p}
\end{equation}%
and%
\begin{eqnarray}
\widetilde{H}_{n_{1},n_{2},n_{3},n_{4}}\left( \mathbf{q}\right)
&=&H_{n_{1},n_{2},n_{3},n_{4}}\left( \mathbf{q}\right) e^{-qd},
\label{hart2} \\
\widetilde{X}_{n_{1},n_{2},n_{3},n_{4}}\left( \mathbf{q}\right) &=&\int 
\frac{d\mathbf{p}\ell ^{2}}{2\pi }\widetilde{H}_{n_{1},n_{2},n_{3},n_{4}}%
\left( \mathbf{p}\right) e^{i\mathbf{q}\times \mathbf{p}\ell ^{2}},
\label{fock2}
\end{eqnarray}%
where $d=3.337$ \AA\ is the separation between the two graphene layers of
the bilayer. The form factors which appear here,%
\begin{eqnarray}
K_{0,0}\left( \mathbf{q}\right) &=&\exp \left( \frac{-q^{2}\ell ^{2}}{4}%
\right) ,  \label{K1} \\
K_{1,1}\left( \mathbf{q}\right) &=&\exp \left( \frac{-q^{2}\ell ^{2}}{4}%
\right) \left( 1-\frac{q^{2}\ell ^{2}}{2}\right) ,  \label{K2} \\
K_{1,0}\left( \mathbf{q}\right) &=&\left( \frac{\left( q_{y}+iq_{x}\right)
\ell }{\sqrt{2}}\right) \exp \left( \frac{-q^{2}\ell ^{2}}{4}\right) ,
\label{K3} \\
K_{0,1}\left( \mathbf{q}\right) &=&\left( \frac{\left( -q_{y}+iq_{x}\right)
\ell }{\sqrt{2}}\right) \exp \left( \frac{-q^{2}\ell ^{2}}{4}\right) ,
\label{K4}
\end{eqnarray}%
capture the character of the two different orbital states. Detailed
expressions for the Hartree and Fock interactions can be found in Appendix~A
of Ref.~\onlinecite{cote1}.

The average values of the density operators $\left\langle \rho
_{n_{1},n_{2}}^{a,b}\left( \mathbf{q}\right) \right\rangle $ are found by
solving the Hartree-Fock equation of motion for the matrix Green's function%
\begin{eqnarray}
G_{n_{1},n_{2}}^{a,b}\left( \mathbf{q,}\tau \right) &=&\frac{1}{N_{\varphi }}%
\sum_{X,X^{\prime }}e^{-\frac{i}{2}q_{x}\left( X+X^{\prime }\right) } \\
&&\times \delta _{X,X^{\prime }-q_{y}\ell ^{2}}G_{n_{1},n_{2}}^{a,b}\left(
X,X^{\prime },\tau \right) ,  \notag
\end{eqnarray}%
with 
\begin{equation}
G_{n_{1},n_{2}}^{a,b}\left( X,X^{\prime },\tau \right) =-\left\langle
Tc_{a,X,n_{1}}\left( \tau \right) c_{b,X^{\prime },n_{2}}^{\dagger }\left(
0\right) \right\rangle .  \label{gmatsubara}
\end{equation}%
When $\tau =0^{-},$ 
\begin{equation}
G_{n_{1},n_{2}}^{a,b}\left( \mathbf{q,}\tau =0^{-}\right) =\left\langle \rho
_{n_{2},n_{1}}^{b,a}\left( \mathbf{q}\right) \right\rangle .  \label{g0}
\end{equation}

The Hartre-Fock equation of motion for this single particle Green's function
is given by%
\begin{eqnarray}
&&\left[ \hslash i\omega _{n}-\left( E_{a,n}-\mu \right) \right]
G_{n,n^{\prime }}^{a,b}\left( \mathbf{q},\omega _{n}\right) =\hslash \delta
_{\mathbf{q},0}\delta _{n,n^{\prime }}\delta _{a,b}  \label{HFmotion} \\
&&+\sum_{c,n_{4}}\overline{\sum_{\mathbf{q}^{\prime }}}U_{c,a}^{H}\left(
n,n_{4},\mathbf{q-q}^{\prime }\right) e^{-i\mathbf{q}\times \mathbf{q}%
^{\prime }\ell ^{2}/2}G_{n_{4},n^{\prime }}^{a,b}\left( \mathbf{q}^{\prime }%
\mathbf{,}\omega _{n}\right)  \notag \\
&&-\sum_{c,n_{4}}\sum_{\mathbf{q}^{\prime }}U_{c,a}^{F}\left( n,n_{4},%
\mathbf{q-q}^{\prime }\right) e^{-i\mathbf{q}\times \mathbf{q}^{\prime }\ell
^{2}/2}G_{n_{4},n^{\prime }}^{c,b}\left( \mathbf{q}^{\prime },\omega
_{n}\right) ,  \notag
\end{eqnarray}%
with the Hartree and Fock potentials 
\begin{equation}
U_{c,a}^{H}\left( n,n_{4},\mathbf{q}\right) =\sum_{n_{1},n_{2}}H_{c,a}\left(
n_{1},n_{2},n,n_{4};-\mathbf{q}\right) \left\langle \rho
_{n_{1},n_{2}}^{c,c}\left( \mathbf{q}\right) \right\rangle ,
\end{equation}%
\begin{equation}
U_{c,a}^{F}\left( n,n_{4},\mathbf{q}\right) =\sum_{n_{1},n_{2}}X_{c,a}\left(
n_{1},n_{4},n,n_{2};-\mathbf{q}\right) \left\langle \rho
_{n_{1},n_{2}}^{c,a}\left( \mathbf{q}\right) \right\rangle .
\end{equation}%
In these equations, $H_{c=a}=H$, $H_{c\neq a}=\widetilde{H}$ and similarly
for $X_{c,a}.$

Equation~(\ref{HFmotion}) constitutes a set of self-consistent equations that
can be solved numerically using the procedure described in Ref.~ \onlinecite{methode}. 
We search amongst the many solutions of this equation
for the one that minimizes the Hartree-Fock energy per electron%
\begin{eqnarray}
\frac{E_{HF}}{N_{0}} &=&\frac{1}{\widetilde{\nu }}\sum_{a,n} E%
_{a,n}\left\langle \rho _{n,n}^{a,a}\left( 0\right) \right\rangle
\label{energy} \\
&&+\frac{1}{2\widetilde{\nu }}\sum_{a,b}\sum_{n_{1},...,n_{4}}\overline{%
\sum_{\mathbf{q}}}H_{a,b}\left( n_{1},n_{2},n_{3},n_{4};\mathbf{q}\right) 
\notag \\
&&\times \left\langle \rho _{n_{1},n_{2}}^{a,a}\left( -\mathbf{q}\right)
\right\rangle \left\langle \rho _{n_{3},n_{4}}^{b,b}\left( \mathbf{q}\right)
\right\rangle  \notag \\
&&-\frac{1}{2\widetilde{\nu }}\sum_{a,b}\sum_{n_{1},...,n_{4}}\sum_{\mathbf{q%
}}X_{a,b}\left( n_{1},n_{4},n_{3},n_{2};\mathbf{q}\right)  \notag \\
&&\times \left\langle \rho _{n_{1},n_{2}}^{a,b}\left( -\mathbf{q}\right)
\right\rangle \left\langle \rho _{n_{3},n_{4}}^{b,a}\left( \mathbf{q}\right)
\right\rangle ,  \notag
\end{eqnarray}%
where $N_{0}$ is the number of electrons in $N=0$.

\section{ORDER PARAMETERS AND PSEUDOSPIN DESCRIPTION}

The set of parameters $\left\{ \left\langle \rho _{n_{1},n_{2}}^{a,b}\left( 
\mathbf{q}\right) \right\rangle \right\} $ fully characterizes a particular
ground state. In the uniform states studied in Refs.~ \onlinecite{yafis1,yafis2,cote1}, 
these parameters were nonzero for $%
\mathbf{q}=0$ only. In this paper, however, we study crystal states
occurring at non-integer filling factor $\widetilde{\nu }$ where we expect a
finite fraction of the electrons, usually $\widetilde{\nu }-\lfloor 
\widetilde{\nu }\rfloor ,$ to crystallize. The set of parameters $\left\{
\left\langle \rho _{n_{1},n_{2}}^{a,b}\left( \mathbf{q}\right) \right\rangle
\right\} $ are then nonzero for $\mathbf{q}=\mathbf{G}$ where $\mathbf{G}$
is a reciprocal lattice vector of the crystal lattice considered.

Generally speaking, crystalline states should occur universally near integer
filling factors in order to maximize the correlations among the
lowest-energy elementary charged excitations. When the charged objects
become more dense at larger departures from integer filling factor they will
begin to overlap. Eventually the various exotic crystalline states will
quantum melt and the electrons will form a fluid state. The methods employed
in this paper are not able to predict the stability range of the crystalline
states.

In Fourier space, the real electronic density in valley (or layer) $a$ is
given by%
\begin{equation}
\left\langle n_{a}\left( \mathbf{G}\right) \right\rangle
=\sum_{n,m=0}^{1}N_{\varphi }K_{n,m}\left( -\mathbf{G}\right) \left\langle
\rho _{n,m}^{a,a}\left( \mathbf{G}\right) \right\rangle .  \label{densi1}
\end{equation}%
We will refer to the inverse Fourier transform $n_{a}\left( \mathbf{r}%
\right) $ of $\left\langle n_{a}\left( \mathbf{G}\right) \right\rangle $ as
the density in the \textit{real space representation} (RSR). We also make
use of another expression for the density:%
\begin{equation}
\left\langle \widetilde{n}_{a}\left( \mathbf{G}\right) \right\rangle
=\sum_{n=0}^{1}\left\langle \rho _{n,n}^{a,a}\left( \mathbf{G}\right)
\right\rangle .  \label{guiding}
\end{equation}%
We refer to the inverse Fourier transform $\widetilde{n}_{a}\left( \mathbf{r}%
\right) $ of $\left\langle \widetilde{n}_{a}\left( \mathbf{G}\right)
\right\rangle $ as the density in the \textit{guiding-center representation}
(GCR). By definition, $\left\langle \widetilde{n}_{a}\left( \mathbf{G}%
=0\right) \right\rangle =\widetilde{\nu }_{a}$ is just the filling factor in
valley $a.$ The form factors $K_{n,m}\left( \mathbf{G}\right) $ are not
taken into account in the GCR so that the character of the different
orbitals $n=0,1$ is lost in the corresponding density.

Interlayer coherence implies that $\left\langle \rho _{n,n}^{a,b\neq
a}\left( \mathbf{G}\right) \right\rangle \neq 0$ while inter-orbital
coherence implies that $\left\langle \rho _{n,m\neq n}^{a,a}\left( \mathbf{G}%
\right) \right\rangle \neq 0.$ In the most general case, both interlayer and
inter-orbital coherences are present and $\left\langle \rho _{n,m\neq
n}^{a,b\neq a}\left( \mathbf{G}\right) \right\rangle \neq 0.$ The different
phases are best described by using a pseudospin language. For the orbital
pseudospin, $\mathbf{S},$ we associate the up state with the $n=0$ orbital
and the down state with the $n=1$ orbital so that in valley $a:$

\begin{eqnarray}
\widetilde{S}_{a,z}\left( \mathbf{G}\right) &=&\frac{1}{2}\left[
\left\langle \rho _{0,0}^{a,a}\left( \mathbf{G}\right) \right\rangle
-\left\langle \rho _{1,1}^{a,a}\left( \mathbf{G}\right) \right\rangle \right]
,  \label{s_4} \\
\widetilde{\mathbf{S}}_{a,\bot }\left( \mathbf{G}\right) &=&\widetilde{S}%
_{a,x}\widehat{\mathbf{x}}+\widetilde{S}_{a,y}\widehat{\mathbf{y}},
\label{s_5} \\
\widetilde{S}_{a,+}\left( \mathbf{G}\right) &=&\widetilde{S}_{a,x}+i%
\widetilde{S}_{a,y}=\left\langle \rho _{0,1}^{a,a}\left( \mathbf{G}\right)
\right\rangle .  \label{s_6}
\end{eqnarray}%
For the interlayer pseudospin, $\mathbf{P}$, we associate the up state with
the $K$ layer and the down state with the $K^{\prime }$ layer so that for
orbital $n$, we have 
\begin{eqnarray}
\widetilde{P}_{n,z}\left( \mathbf{G}\right) &=&\frac{1}{2}\left[
\left\langle \rho _{n,n}^{K,K}\left( \mathbf{G}\right) \right\rangle
-\left\langle \rho _{n,n}^{K^{\prime },K^{\prime }}\left( \mathbf{G}\right)
\right\rangle \right] ,  \label{s_7} \\
\widetilde{\mathbf{P}}_{\bot ,n}\left( \mathbf{G}\right) &=&\widetilde{P}%
_{n,x}\widehat{\mathbf{x}}+\widetilde{P}_{n,y}\widehat{\mathbf{y}},
\label{s_8} \\
\widetilde{P}_{n,+}\left( \mathbf{G}\right) &=&\widetilde{P}_{n,x}+i%
\widetilde{P}_{n,y}=\left\langle \rho _{n,n}^{K,K^{\prime }}\left( \mathbf{G}%
\right) \right\rangle .  \label{s_9}
\end{eqnarray}%
Note that these fields are defined in the GCR. To get them in the RSR, we
multiply each $\left\langle \rho _{n,m}^{a,a}\left( \mathbf{G}\right)
\right\rangle $ in these definitions with $N_{\varphi }K_{n,m}\left( -%
\mathbf{G}\right) $. From now on, we use the notation $\widetilde{\mathbf{S}}%
,\widetilde{\mathbf{P}},\widetilde{n}$ to refer to the fields in the GCR and
the notation $\mathbf{S},\mathbf{P},n$ for the fields in the RSR. The two
views give separate interesting insights into the nature of the crystal
states, but the RSR is more closely related to experimental probes like a
scanning tunneling microscope (STM).

An exact description of the state of one electron in $N=0$ is given by the
four complex components of the spinor $\left( c_{K,X,0}^{\dagger
},c_{K,X,1}^{\dagger },c_{K^{\prime },X,0}^{\dagger },c_{K^{\prime
},X,1}^{\dagger }\right) .$ Note that, in order to limit the range of
possible states, we will restrict our attention to circumstances in which
the $N=0$ states are maximally polarized. As we mentioned in Sec.~II, we
expect that partially polarized states will be common at large interlayer
potentials. The ideas explained here are readily generalized to include this
possibility. For a CP$^{3}$ spinor, the norm and the absolute phase are
fixed so that a given electronic state is defined by $6$ independent
components or angles\cite{gosh}. The $12$ classical fields $\widetilde{%
\mathbf{P}}_{n=0,1}\left( \mathbf{G}\right) ,\widetilde{\mathbf{S}}_{\pm
K}\left( \mathbf{G}\right) $ that we defined in Eqs.~(\ref{s_4}-\ref{s_9})
have a simple physical interpretation but they do not provide a full
description of a given phase. Moreover, these fields are not
independent variables and their norm is not fixed. Both the modulus and the
orientation of these pseudospins may vary in space. In fact, the sum rule%
\begin{equation}
\sum_{a,b}\sum_{n,m}\sum_{\mathbf{G}}\left\vert \left\langle \rho
_{n,m}^{a,b}\left( \mathbf{G}\right) \right\rangle \right\vert ^{2}=%
\widetilde{\nu },
\end{equation}%
that applies when the many-electron state is approximated by a single Slater
determinant becomes, in pseudospin language, 
\begin{eqnarray}
&&\sum_{\mathbf{G}}\left[ \frac{1}{4}\left\vert \rho _{K}\left( \mathbf{G}%
\right) +\rho _{K^{\prime }}\left( \mathbf{G}\right) \right\vert ^{2}\right.
\\
&&-\left\vert \widetilde{P}_{z,0}\left( \mathbf{G}\right) -\widetilde{P}%
_{z,1}\left( -\mathbf{G}\right) \right\vert ^{2}  \notag \\
&&+2\left\vert \widetilde{S}_{K}\left( \mathbf{G}\right) \right\vert
^{2}+2\left\vert \widetilde{S}_{K^{\prime }}\left( \mathbf{G}\right)
\right\vert ^{2}  \notag \\
&&+2\left\vert \widetilde{\mathbf{P}}_{0}\left( \mathbf{G}\right)
\right\vert ^{2}+2\left\vert \widetilde{\mathbf{P}}_{1}\left( \mathbf{G}%
\right) \right\vert ^{2}  \notag \\
&&\left. +2\left\vert \left\langle \rho _{0,1}^{K,K^{\prime }}\left( \mathbf{%
G}\right) \right\rangle \right\vert ^{2}+2\left\vert \left\langle \rho
_{1,0}^{K,K^{\prime }}\left( \mathbf{G}\right) \right\rangle \right\vert ^{2}%
\right]  \notag \\
&=&\widetilde{\nu. }  \notag
\end{eqnarray}, where we have defined $\rho_a=\rho_{0,0}^{a,a}$. 

Skyrmion crystals with intervalley pseudospin textures have been studied
extensively in semiconductor 2DEG as well as in graphene monolayers\cite%
{cotecp3,joglekar,jobidon}. In bilayer graphene, we have the additional possibility
of orbital pseudospin texture. This type of texture is particularly
interesting because it gives rise to textures of electric dipoles in the
plane of the layer. As shown in Refs.~\onlinecite{cote1,shizuya}, the
coupling of the 2DEG with an electric field $\mathbf{E}\left( \mathbf{r}%
\right) =-\nabla \phi \left( \mathbf{r}\right) $ in the plane of the layers
can be written as $H_{ext}=-e\int d\mathbf{r}\;n\left( \mathbf{r}\right)
\phi \left( \mathbf{r}\right) $ where $n\left( \mathbf{r}\right) $ is the
Fourier transform of the total density $n_{K}\left( \mathbf{G}\right)
+n_{K^{\prime }}\left( \mathbf{G}\right) $ (see Eq.~(\ref{densi1}). With the
form factor defined in Eqs.~(\ref{K1}-\ref{K4}) and $j=K,K^{\prime}$, 
\begin{eqnarray}
H_{ext} &=&-\frac{eN_{\varphi }}{S}\sum_{j}\sum_{\mathbf{G}%
}\left[ \left( 1-\frac{G^{2}\ell ^{2}}{4}%
\right) \overline{\rho }_{j}\left( -\mathbf{G}\right) \right. \\
&&+\left( \frac{G^{2}\ell ^{2}}{2}\right) \overline{\rho }_{j,z}\left( -%
\mathbf{G}\right)  \notag \\
&&\left. -\sqrt{2}i\left( G_{x}\ell \;\overline{\rho }_{j,x}\left( -\mathbf{G%
}\right) -G_{y}\ell \;\overline{\rho }_{j,y}\left( -\mathbf{G}\right)
\right) \right] \phi \left( \mathbf{G}\right) ,  \notag
\end{eqnarray}%
where we have defined $\overline{\rho }_{j}\left( \mathbf{G}\right) =\exp
\left( -G^{2}\ell ^{2}/4\right) \rho _{j}\left( \mathbf{G}\right) $. In real
space,

\begin{eqnarray}
H_{ext} &=&-eN_{\varphi }\sum_{j}\int d\mathbf{r}\left[ 
\overline{\rho }_{j}\left( \mathbf{r}\right) \phi \left( \mathbf{r}\right)
\right.  \notag \\
&&-\left. \frac{1}{4}\left( \overline{\rho }_{j}\left( \mathbf{r}\right)
\ell ^{2}-2\overline{\rho }_{j,z}\left( \mathbf{r}\right) \ell ^{2}\right)
\left( \mathbf{\nabla }\cdot \mathbf{E}\left( \mathbf{r}\right) \right) %
\right]  \label{edipolep} \\
&&+\sqrt{2}\ell eN_{\varphi }\int d\mathbf{r}\left[ \overline{\rho }%
_{j,x}\left( \mathbf{r}\right) E_{x}\left( \mathbf{r}\right) -\overline{\rho 
}_{j,y}\left( \mathbf{r}\right) E_{y}\left( \mathbf{r}\right) \right] , 
\notag
\end{eqnarray}%
so that we can identify 
\begin{eqnarray}
\mathbf{d}_{a}\left( \mathbf{G}\right) &=&-e\sqrt{2}\ell N_{\varphi
}e^{-G^{2}\ell ^{2}/4}  \label{dipole} \\
&&\times \left( \left\langle \rho _{a,x}\left( \mathbf{G}\right)
\right\rangle \widehat{\mathbf{x}}-\left\langle \rho _{a,y}\left( \mathbf{G}%
\right) \right\rangle \widehat{\mathbf{y}}\right)  \notag
\end{eqnarray}%
with the Fourier transform of an electric dipole field in layer $a.$ The
orientation of the dipole vector at each point in space is simply related to
the orientation of the orbital pseudospin vector. It follows that
crystals with orbital pseudospin textures then have electric-dipole
textures. Orbital Skyrmion crystals are the electric analog of spin
Skyrmions crystals in which it is the magnetization that varies in space.
Note that in a uniform electric field, the second term in Eq.~(\ref{edipolep}%
) is zero and $H_{ext}$ gives the monopole and dipole terms of the
interaction energy of the electrons with the external electric field.

\section{ISOLATED SKYRMIONS}

\begin{figure}[tbph]
\includegraphics[scale=1]{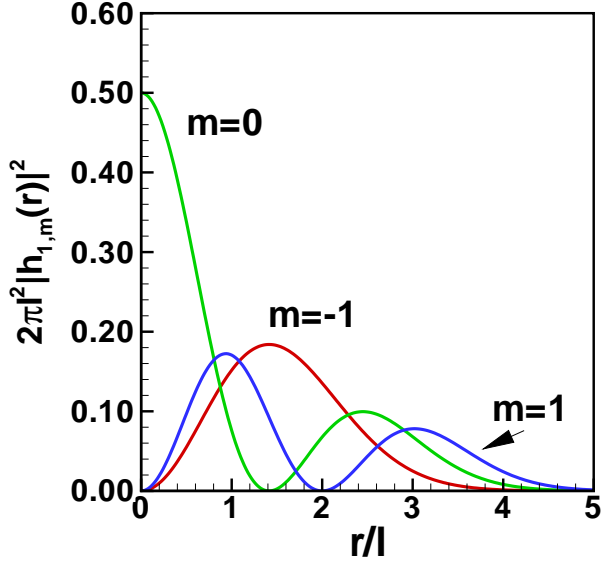}
\caption{(Color online) Density profile for the wave functions in $n=1$ with 
$m=-1,0,1.$}
\label{fig3}
\end{figure}

Before we can analyze the results of the numerical calculations for the
Skyrmion crystal states, we need to know the density and pseudospin patterns
associated with a single orbital Skyrmion located at $\mathbf{r}=0.$ We use,
in this section, the symmetric gauge which is more convenient for this
problem. We take $\mathbf{A}=\left( -By/2,Bx/2,0\right) $ for the vector
potential. The eigenfunctions of the kinetic Hamiltonian $H=\left( \mathbf{p}%
+e\mathbf{A/}c\right) ^{2}/2m_0$ (where $-e$ is the charge of an
electron and $m_0$ the electronic mass) are given by 
\begin{equation}
h_{n=0,m}\left( \mathbf{r}\right) =\frac{1}{\sqrt{2\pi 2^{m}m!}\ell }\left( 
\frac{r}{\ell }\right) ^{m}e^{-im\varphi }e^{-r^{2}/4\ell ^{2}},
\end{equation}%
with $m=0,1,2,...$ for Landau level $n=0$ and by 
\begin{eqnarray}
h_{n=1,m}\left( \mathbf{r}\right) &=&\frac{1}{\sqrt{\pi 2^{\left\vert
m\right\vert +1}(m+1)!}}\frac{1}{\ell }\left( \frac{r}{\ell }\right)
^{\left\vert m\right\vert } \\
&&\times e^{-im\varphi }e^{-r^{2}/4\ell ^{2}}L_{1+\left( \frac{m-\left\vert
m\right\vert }{2}\right) }^{\left\vert m\right\vert }\left( \frac{r^{2}}{%
2\ell ^{2}}\right) ,  \notag
\end{eqnarray}%
with $m=-1,0,1,2,...$for Landau level $n=1$ and $L_{n}^{m}\left( x\right) $
is a generalized Laguerre polynomial. (Note that $h_{n,m}$ is a different function
than $h_{n,X}$ introduced previously). Figure~\ref{fig3} shows the density
profile $n_{1}\left( \mathbf{r}\right) =\left\vert h_{1,m}\left( \mathbf{r}%
\right) \right\vert ^{2}$ for the eigenstates with $m=-1,0,1.$

The index $m\geq -n$ gives the angular momentum i.e. 
\begin{equation}
L_{z}h_{n,m}\left( \mathbf{r}\right) =-\hslash mh_{n,m}\left( \mathbf{r}%
\right) ,
\end{equation}%
while the energy of each state $\left( n,m\right) $ is given by 
\begin{equation}
E_{n}^{0}=\left( n+1/2\right) \hslash \omega _{c},
\end{equation}%
where $\omega _{c}$ is the cyclotron frequency. Note that for a filled
level, we have%
\begin{eqnarray}
n_{0}\left( \mathbf{r}\right) &=&\sum_{m=0}^{\infty }\left\vert
h_{0,m}\left( \mathbf{r}\right) \right\vert ^{2}=\frac{1}{2\pi \ell ^{2}},
\label{rule1} \\
n_{1}\left( \mathbf{r}\right) &=&\sum_{m=-1}^{\infty }\left\vert
h_{1,m}\left( \mathbf{r}\right) \right\vert ^{2}=\frac{1}{2\pi \ell ^{2}}.
\label{rule2}
\end{eqnarray}

\subsection{Spin Skyrmion}

In a semiconductor 2DEG, the state $\left\vert \mathcal{S}\right\rangle $
corresponding to the addition of one spin Skyrmion with topological charge $%
Q=1$ at $\mathbf{r}=0$ to the spin-polarized ground state $\left\vert
GS\right\rangle =\prod\limits_{m=0}^{\infty }c_{m,\,\uparrow }^{\dag
}\left\vert 0\right\rangle $ at $\nu =1$ (Landau level $N=0$ is implicitly
assumed) is given by 
\begin{equation}
\left\vert \mathcal{S}\right\rangle =\prod\limits_{m=0}^{\infty }\left[
-u_{m}c_{\downarrow ,m+1}^{\dag }+v_{m}c_{\uparrow ,m}^{\dag }\right]
c_{\downarrow ,0}^{\dag }\left\vert 0\right\rangle ,
\end{equation}%
with the constraint that 
\begin{equation}
\left\vert u_{m}\right\vert ^{2}+\left\vert v_{m}\right\vert ^{2}=1.
\end{equation}%
The anti-Skyrmion state is given by%
\begin{equation}
\left\vert \mathcal{AS}\right\rangle =\prod\limits_{m=0}^{\infty }\left[
u_{m}c_{\downarrow ,m}^{\dag }+v_{m}c_{\uparrow ,m+1}^{\dag }\right]
\left\vert 0\right\rangle .
\end{equation}%
The values of $u_{m}$ and $v_{m}$ depend on details on the Zeeman coupling
and the electron-electron interaction and can be fixed by energy
minimization.

The excitations $\left\vert \mathcal{S}\right\rangle ,\left\vert \mathcal{AS}%
\right\rangle $ have collective coherence between single-particle states
with different spins and angular momenta values that differ by one. The RSR
of the field $\mathbf{S}_{-}\left( \mathbf{r}\right) =S_{x}\left( \mathbf{r}%
\right) -iS_{y}\left( \mathbf{r}\right) $ for the Skyrmion state is given by 
\begin{eqnarray}
S_{-}\left( \mathbf{r}\right) &=&\left\langle \mathcal{S}\right\vert \Psi
_{\downarrow }^{\dag }\left( \mathbf{r}\right) \Psi _{\uparrow }\left( 
\mathbf{r}\right) \left\vert \mathcal{S}\right\rangle \\
&=&-\sum_{m=0}^{\infty }h_{0,m+1}^{\ast }\left( \mathbf{r}\right)
h_{0,m}\left( \mathbf{r}\right) u_{m}^{\ast }v_{m},  \notag
\end{eqnarray}%
where $\Psi _{\sigma }^{\dag }\left( \mathbf{r}\right) $ is the field
operator that creates an electron at $\mathbf{r}$ with spin $\sigma .$ The
corresponding density of the two spin components in the RSR\ are then%
\begin{eqnarray}
n_{\uparrow }\left( \mathbf{r}\right) &=&\left\langle \mathcal{S}\right\vert
\Psi _{\uparrow }^{\dag }\left( \mathbf{r}\right) \Psi _{\uparrow }\left( 
\mathbf{r}\right) \left\vert \mathcal{S}\right\rangle \\
&=&\sum_{m=0}^{\infty }\left\vert h_{0,m}\left( \mathbf{r}\right)
\right\vert ^{2}\left\vert v_{m}\right\vert ^{2},  \notag
\end{eqnarray}%
and%
\begin{eqnarray}
n_{\downarrow }\left( \mathbf{r}\right) &=&\left\langle \mathcal{S}%
\right\vert \Psi _{_{\downarrow }}^{\dag }\left( \mathbf{r}\right) \Psi
_{_{\downarrow }}\left( \mathbf{r}\right) \left\vert \mathcal{S}\right\rangle
\\
&=&\left\vert h_{0,0}\left( \mathbf{r}\right) \right\vert
^{2}+\sum_{m=0}^{\infty }\left\vert h_{0,m+1}\left( \mathbf{r}\right)
\right\vert ^{2}\left\vert u_{m}\right\vert ^{2}.  \notag
\end{eqnarray}%
The total density is of course 
\begin{equation}
n\left( \mathbf{r}\right) =n_{\uparrow }\left( \mathbf{r}\right)
+n_{\downarrow }\left( \mathbf{r}\right) ,
\end{equation}%
while the $z-$component of the spin field is given by%
\begin{equation}
S_{z}\left( \mathbf{r}\right) =\frac{1}{2}\left[ n_{\uparrow }\left( \mathbf{%
r}\right) -n_{\downarrow }\left( \mathbf{r}\right) \right] .
\end{equation}%
(Note that for a filled level, $\left\vert S_{z}\left( \mathbf{r}\right)
\right\vert =1/4\pi \ell ^{2}$.)

\begin{figure}[tbph]
\includegraphics[scale=1]{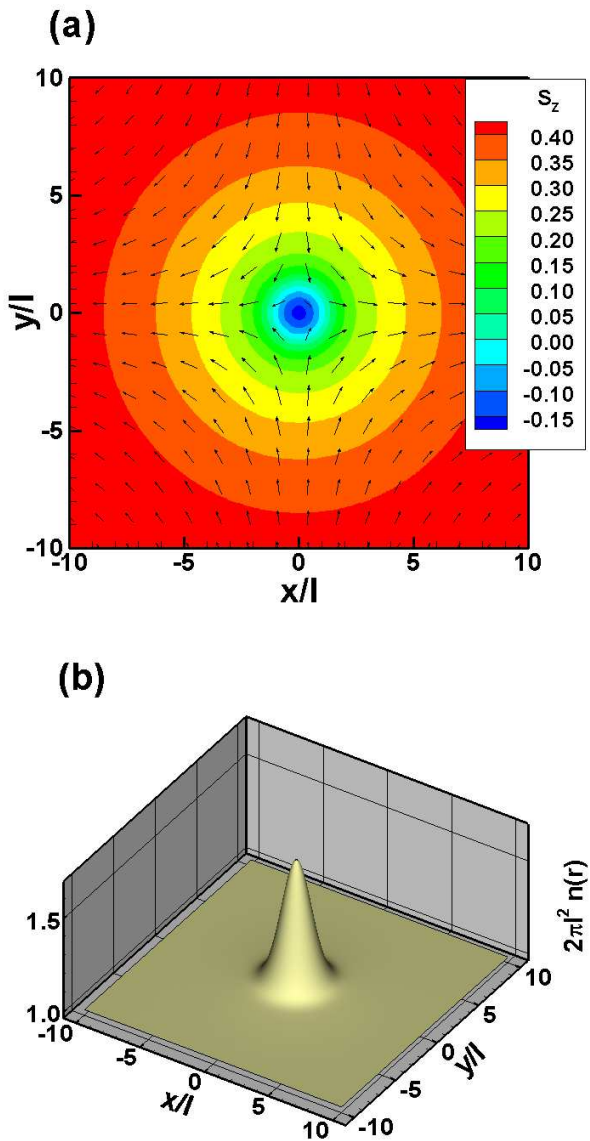}
\caption{(Color online) A charge $Q=1$ Skyrmion for the parameters of Eq.~(%
\protect\ref{umvm}) with $\Delta =4$. (a) Spin profile in $x-y$ plane. (b)
Total density $n\left( \mathbf{r}\right) $. }
\label{fig4}
\end{figure}

Figure~\ref{fig4} shows the density and spin patterns for a spin Skyrmion
with $Q=1$ added to a filled Landau level. We have chosen for this figure
the simple expression 
\begin{equation}
u_{m}=\sqrt{\frac{\Delta }{m+10+\Delta }},v_{m}=\sqrt{\frac{m+10}{%
m+10+\Delta }},  \label{umvm}
\end{equation}%
where $\Delta $ is the Skyrmion size. This expression gives a density and
pseudospin pattern for the Skyrmion which is qualitatively close to that
given by the energy minimization\cite{fertigskyrmion}. Note that all spin
and pseudospin densities in this figure and the subsequent ones in this
paper are in units of $1/2\pi \ell ^{2}.$ This corresponds to the density of
a filled Landau level (see Eqs.~(\ref{rule1},\ref{rule2})).

\subsection{Orbital Skyrmion}

We now consider the case where all electrons have spin up and we try to make
an orbital Skyrmion by flipping some orbital pseudospins from $n=0$ to $n=1.$%
\ For simplicity, we assume full valley and spin polarization so that we can
drop the layer and spin indices. Full valley polarization is expected at all
odd integer filling factors when the interlayer potential is strong. The
ground state at $\widetilde{\nu }=1$ is given by 
\begin{equation}
\left\vert GS\right\rangle =\prod\limits_{m=0}^{\infty }c_{0,m}^{\dag
}\left\vert 0\right\rangle .
\end{equation}%
An orbital anti-Skyrmion state can be written as%
\begin{equation}
\left\vert \mathcal{AS}\right\rangle =\prod\limits_{m=-1}^{\infty }\left[
u_{m}c_{1,m}^{\dag }+v_{m}c_{0,m+2}^{\dag }\right] \left\vert 0\right\rangle
.
\end{equation}%
We see that the angular momentum difference $\Delta m=m_{1}-m_{0}=-2$
because the lowest value of $m$ in $n=1$ is $m=-1.$

For the Skyrmion excitation with the same vorticity, we have three choices
corresponding to $p=-1,0,1$ in the expression

\begin{equation}
\left\vert \mathcal{S}_{p}\right\rangle =\prod\limits_{m=0}^{\infty }\left[
-u_{m}c_{1,m+2}^{\dag }+v_{m}c_{0,m}^{\dag }\right] c_{1,p}^{\dag
}\left\vert 0\right\rangle .  \label{skyrmion}
\end{equation}%
The RSR of the pseudospin field $\mathbf{S}_{-}\left( \mathbf{r}\right) $ is
given by 
\begin{equation}
S_{-}\left( \mathbf{r}\right) =-\sum_{m=0}^{\infty }h_{1,m+2}^{\ast }\left( 
\mathbf{r}\right) h_{0,m}\left( \mathbf{r}\right) u_{m}^{\ast }v_{m},
\label{spin1}
\end{equation}%
while the densities in $n=0$ and $n=1$ are given by%
\begin{eqnarray}
n_{0}\left( \mathbf{r}\right) &=&\left\langle \mathcal{S}_{p}\right\vert
\Psi _{0}^{\dag }\left( \mathbf{r}\right) \Psi _{0}\left( \mathbf{r}\right)
\left\vert \mathcal{S}_{p}\right\rangle \\
&=&\sum_{m=0}^{\infty }\left\vert h_{0,m}\left( \mathbf{r}\right)
\right\vert ^{2}\left\vert v_{m}\right\vert ^{2},  \notag
\end{eqnarray}%
and 
\begin{eqnarray}
n_{1,p}\left( \mathbf{r}\right) &=&\left\langle \mathcal{S}_{p}\right\vert
\Psi _{1}^{\dag }\left( \mathbf{r}\right) \Psi _{1}\left( \mathbf{r}\right)
\left\vert \mathcal{S}_{p}\right\rangle \\
&=&\left\vert \phi _{1,p}\left( \mathbf{r}\right) \right\vert
^{2}+\sum_{m=0}^{\infty }\left\vert h_{1,m+2}\left( \mathbf{r}\right)
\right\vert ^{2}\left\vert v_{m}\right\vert ^{2}.  \notag
\end{eqnarray}%
For the orbital Skyrmion, the \textit{total} density is given by 
\begin{eqnarray}
n\left( \mathbf{r}\right) &=&\sum_{i,j=0}^{1}\left\langle \mathcal{S}%
_{p}\right\vert \Psi _{i}^{\dag }\left( \mathbf{r}\right) \Psi _{j}\left( 
\mathbf{r}\right) \left\vert \mathcal{S}_{p}\right\rangle  \label{ntotal} \\
&=&n_{0}\left( \mathbf{r}\right) +n_{1,p}\left( \mathbf{r}\right) +2\mathrm{Re}%
\left[ S_{-}\left( \mathbf{r}\right) \right]  \notag
\end{eqnarray}%
and includes the extra contribution $2\mathrm{Re}\left[ S_{-}\left( \mathbf{r}%
\right) \right] .$

For the $z-$component 
\begin{equation}
S_{z}\left( \mathbf{r}\right) =\frac{1}{2}\left[ n_{0}\left( \mathbf{r}%
\right) -n_{1,p}\left( \mathbf{r}\right) \right] .
\end{equation}

\begin{figure}[tbph]
\includegraphics[scale=1]{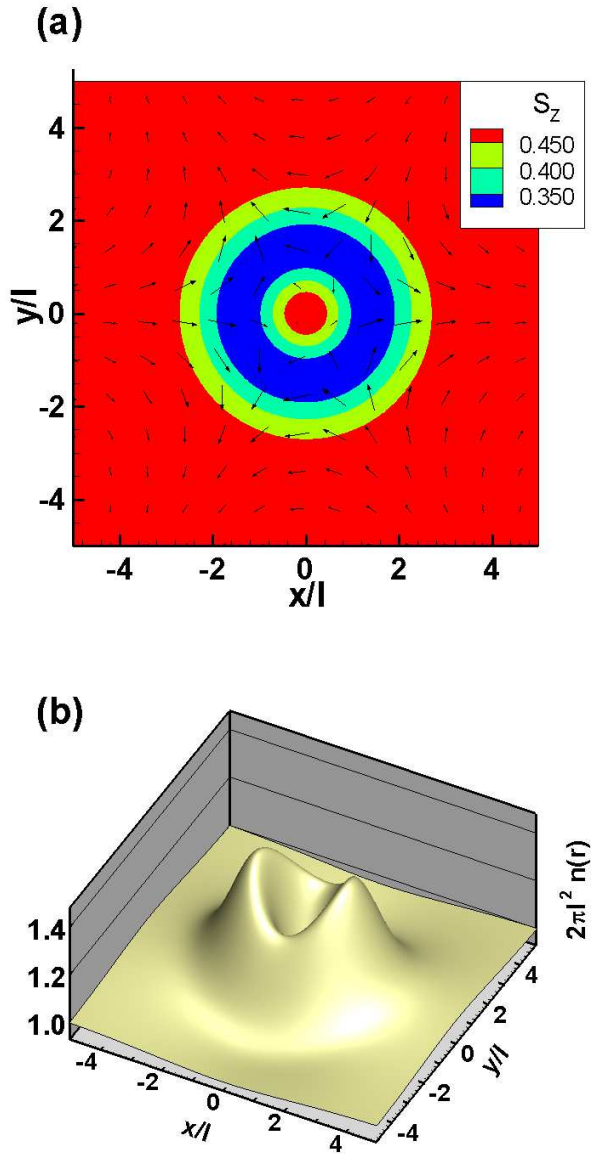}
\caption{(Color online) An orbital Skyrmion with $p=-1$ added to a filled
Landau level for the parameters of Eq.~(\protect\ref{umvm}) with $\Delta
=0.05$. (a) Orbital pseudospin profile in the RSR. (b) Total density $%
n\left( \mathbf{r}\right) $ in the RSR.}
\label{fig5}
\end{figure}

To give an example, we take $p=-1$ and choose again Eq.~(\ref{umvm}) with $%
\Delta =0.05.$ The total density and pseudospin patterns are represented in
Fig.~\ref{fig5}. These patterns differ markedly from those of a spin
Skyrmion. The coupling $m_{1}-m_{0}=-2$ between the angular momenta in $n=0$
and $n=1$ makes the orbital-pseudospin vector to rotate by $4\pi $ instead
of $2\pi $ around the Skyrmion center. Because the value of $u_{m}$ is
small, the profile of the $z$-component of $S_{z}\left( \mathbf{r}\right) $
is basically the (inverted) density profile $n_{1}\left( \mathbf{r}\right) $
(see Fig.~\ref{fig3}). Also, the total density $n\left( \mathbf{r}\right) $
is anisotropic. Comparing this profile with that of $N_{p}\left( \mathbf{r}%
\right) \equiv n_{0}\left( \mathbf{r}\right) +n_{1,p}\left( \mathbf{r}%
\right) $ (not shown in the figure) which is isotropic, we understand that
the anisotropy comes from the term $2\mathrm{Re}\left[ S_{-}\left( \mathbf{r}%
\right) \right] $ in Eq.~(\ref{ntotal}). The density profile is typical of
the density pattern of an electron in state $n=0,p=-1$ (see Fig.~\ref{fig3}).

Our choice for the $u_{m},v_{m}$ in this example is motivated by the fact
that it reproduces qualitatively the density we observed in the crystal
phases that we discuss later. Of course, the correct values of $u_{m},v_{m}$
must be obtained by energy minimization i.e. by solving the eigenvalue
equation for the Hartree-Fock Hamiltonian of an isolated Skyrmion. This is
discussed in Ref.~\onlinecite{fertigskyrmion}. For the same reason, we
choose in Eq.~(\ref{skyrmion}) a pairing with $\Delta m=-2$ because the
vorticity of the Skyrmions we get in our crystals is $4\pi .$

It is clear from Eq.~(\ref{skyrmion}) that there are many variations of the
microscopic wavefunction $\left\vert \mathcal{S}_{p}\right\rangle $ that we
can make that would lead to pseudospin textures with different topological
and real electric charges. A study of these different solutions is, however,
beyond the scope of this paper.

\subsection{Skyrmion energy}

Orbital Skyrmions could occur, for example, at $\nu =-3$ when the bias is
strong enough for all the charge to be transferred to the state $\left\vert
K^{\prime },0\right\rangle .$ When $\zeta _{1}=0$, this state occurs\cite%
{yafis2} for $\Delta _{B}\geq 0.0012e^{2}/\kappa \ell .$ In this state, the
gap between the $n=0$ and $n=1$ state, $\beta \Delta _{B}$, is very small
and one would expect orbital Skyrmions, in analogy with spin Skyrmions, to
have lower energy than the corresponding electron quasiparticles. The
energetics of an orbital Skyrmion is however different from its spin analog
as we now show.

With all electrons in the state $\left\vert K^{\prime },0\right\rangle $,
the energy per electron is given by 
\begin{eqnarray}
\frac{E}{N_{0}} &=&-\frac{1}{2}\Delta _{B}-\frac{1}{2}X_{0,0,0,0}\left(
0\right) \\
&=&-\frac{1}{2}\Delta _{B}-\frac{1}{2}\sqrt{\frac{\pi }{2}}.  \notag
\end{eqnarray}%
The energy needed to remove one electron in orbital $m$ from the ground
state (i.e. the energy to create one hole) is given by%
\begin{equation}
E_{hole}=\frac{1}{2}\Delta _{B}+X_{0,0,0,0}\left( 0\right) ,
\end{equation}%
while the energy required to add one electron in $n=1$ is given by%
\begin{eqnarray}
E_{e} &=&-\frac{1}{2}\Delta _{B}+\beta \Delta _{B}-X_{0,1,1,0}\left( 0\right)
\\
&=&-\frac{1}{2}\Delta _{B}+\beta \Delta _{B}-\frac{1}{2}\sqrt{\frac{\pi }{2}}
\notag
\end{eqnarray}%
and is negative at zero bias. The energy to create an electron (in $n=1)$
and hole (with $n=0)$ pair with infinite separation is thus 
\begin{equation}
\Delta _{eh,(orbital)}=E_{e}+E_{hole}=\beta \Delta _{B}+\frac{1}{2}\sqrt{%
\frac{\pi }{2}}.
\end{equation}%
Let's compare this result with the energy needed to create an electron (in $%
n=0$) and hole (in $n=0$) pair with different spins i.e. 
\begin{equation}
\Delta _{eh,(spin)}=g\mu _{B}B+\sqrt{\frac{\pi }{2}}.
\end{equation}%
The first term on the right-hand side of this last equation is the Zeeman
energy. The energy needed to flip an orbital pseudospin is smaller than the
energy to flip a spin (at zero bias and Zeeman couplings). It follows that
the condition required to excite a Skyrmion pair i.e. $\Delta
_{skyrmion-antiskyrmion}<\Delta _{eh,(orbital)}$ is more restrictive for an
orbital Skyrmion than for a spin Skyrmion. $\Delta _{eh,(orbital)}<\Delta
_{eh,(spin)}$ because of the presence of the extra exchange energy, $%
X_{0,1,1,0}\left( 0\right) ,$ between different orbitals which is not
present in the spin case.

Another important difference between the orbital and spin Skyrmions is that
the exchange energy is smaller in $n=1$ than in $n=0$ i.e. $%
X_{1,1,1,1}\left( 0\right) =\frac{3}{4}X_{0,0,0,0}\left( 0\right) .$ For
this reason, the gain in exchange energy obtained by making an orbital
pseudospin texture is not as big as for a spin texture.

A detailed numerical calculation of the energy of the Skyrmion and
anti-Skyrmion excitations under finite bias thus has to be made in order to
compare their energy with those of the electron and hole excitations. This
can by done by energy minimization, using the method described in Ref.~\onlinecite{fertigskyrmion}. Despite our efforts, we have not been able so
far to achieve sufficient precision with our numerical code to classify
energetically these different solutions. We thus concentrate, in this paper,
on crystal solutions which are a lot easier to compute.

We remark that orbital Skyrmions have been studied previously in a
conventional semiconductor 2DEG\cite{lilly}. They were called, in this
context, inter-Landau-level Skyrmions and involved spin flips between the $%
n=0$ spin down state and the $n=1$ spin up state. Since there is no exchange
energy between states with different spin indices, the energetics of these
inter-Landau-level Skyrmions is different from that of our orbital Skyrmions
in which orbital pseudospin flips occur between states with the same spin. The
conclusion of Ref.~\onlinecite{lilly} that inter-Landau-level Skyrmions are
never the lowest lying charged excitation cannot be applied to our system.

\section{SKYRMION CRYSTALS}

We now consider the ground state of the 2DEG in bilayer graphene at non
integer filling factors $\widetilde{\nu }\in \left[ 1,3\right] $. We look
for crystal solutions of Eq.~(\ref{HFmotion}) allowing for the possibility
of both valley and orbital pseudospin textures. We do not attempt to make an
exhaustive study of the phase diagram of the 2DEG since considering all the
possible crystal states would be a formidable task. We restrict ourselves to
square and triangular lattices with one and two electrons per unit cell and
single out the state with the lowest energy. We take $\zeta _{1}=0$ and
discuss the effect of a finite $\zeta _{1}$ in the conclusion of this paper.

All results presented in this paper are for a magnetic field $B=10$ T. We
measure the energy in units of $e^{2}/\kappa \ell =0.036$ eV for 
$\kappa =5$ appropriate for graphene on a SiO$_{2}$ substrate. For the
validity of the two-band model used in our calculations, we need $\Delta
_{B}<<\gamma _{1}=0.39$ eV i.e. $\Delta _{B}/\left( e^{2}/\kappa \ell
\right) <<11.$

We start by presenting the crystal state at large bias for $\widetilde{\nu }$
around $1$ because the Skyrmion at each site is close to the simple solution
we presented in Fig.~\ref{fig5} in this case. We then study the crystals for 
$\widetilde{\nu }$ near $1$ and $3$. These two filling factors give very
similar solutions. We end with filling factor near $2$ where radically
different solutions are obtained.

\subsection{Orbital Skyrmion crystal at large bias}

\begin{figure}[tbph]
\includegraphics[scale=1]{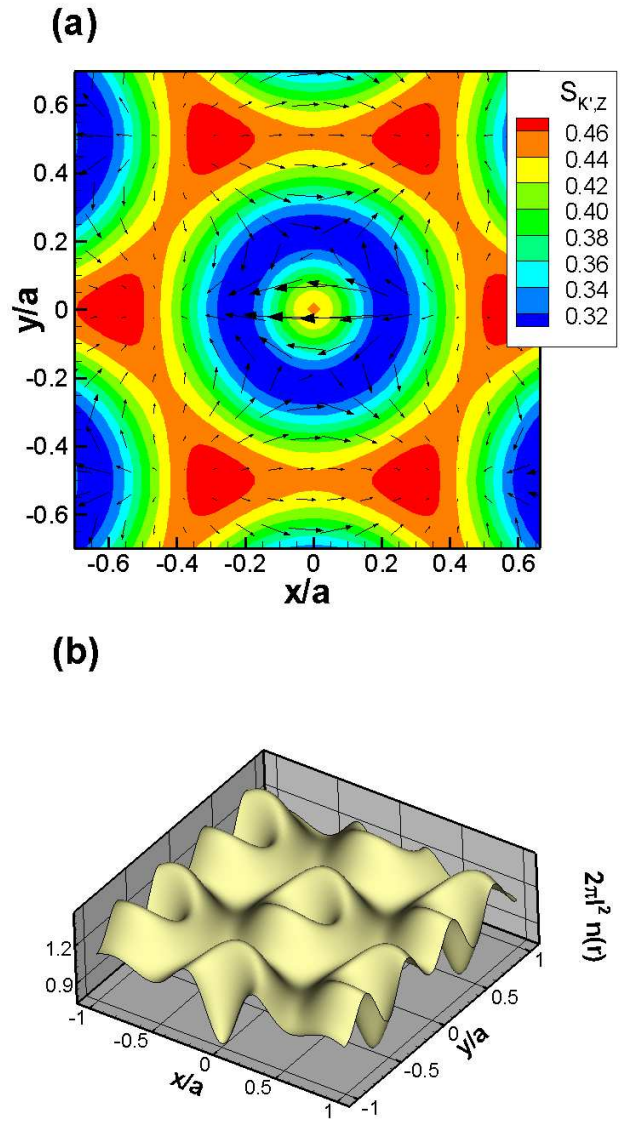}
\caption{(Color online) Orbital Skyrmion crystal at $\widetilde{\protect\nu }%
=1.2$ and $\Delta _{B}/\left( e^{2}/\protect\kappa \ell \right) =1.28.$ (a)
Pseudospin texture in the RSR. (b) Total density $n\left( \mathbf{r}\right) $
in the RSR.}
\label{fig6}
\end{figure}

The simplest crystalline structure occurs at large bias with $\widetilde{\nu 
}$ around $1.$ In this case, valley $K$ is empty and the charge is entirely
in valley $K^{\prime }.$ This corresponds to the situation we studied in
Sec.~V(b). The crystal solution is a triangular lattice of orbital Skyrmions
with one Skyrmion per lattice site. We show an example of this solution for $%
\widetilde{\nu }=1.2,\Delta _{B}/\left( e^{2}/\kappa \ell \right) =1.28$ in
Fig.~\ref{fig6}. We see that for each Skyrmion in the lattice, the
pseudospin and density profiles (in the RSR) are close to the very crude
Skyrmion solution we illustrated in Fig.~\ref{fig5}. Note that if we use the
GCR\ instead, the pseudospin and density profile for each Skyrmion are
exactly those of the usual spin Skyrmion we illustrated in Fig.~\ref{fig4}
i.e. the pseudospins rotates by $2\pi $ around the center of the Skyrmion
and the pseudospins point downward at the center (we show this in Fig.~\ref%
{fig7}).

The Wigner crystal solution (i.e. no orbital pseudospin texture) can be
found if the bias is taken to be extremely large i.e. of the order of $%
\Delta _{B}/\left( e^{2}/\kappa \ell \right) \approx 30$ which is well
beyond the limit of validity of our model. If we compare the interaction
energy of the Skyrme crystal at $\Delta _{B}/\left( e^{2}/\kappa \ell
\right) =1.28$ with that of the Wigner crystal at $\Delta _{B}/\left(
e^{2}/\kappa \ell \right) =30$, we find that the energy of the former is
lower than that of the later by approximately $0.5\%.$ (The interaction
energy includes all terms in Eq.~(\ref{energy}) with the exception of the
bias energy). The Hartree part of the total energy is bigger in the Skyrme
than in the Wigner crystal while the exchange (Fock) energy is more negative
in the Skyrme crystal probably because Skyrmions are larger objects and
overlap more with their neighbors. It could be also that isolated Skyrmions
have lower energy than electron or hole quasiparticles.  As mentioned earlier, we have 
not been able to confirm that Skyrmions have lower energy than 
isolated electrons and holes in the dilute limit using separate isolated quasiparticle calculations. 
If indeed Skrymions are only stable beyond a minimum density, the crystal stability must be 
related to inter-Skyrmion exchange energies

\subsection{Skyrme crystals near $\widetilde{\protect\nu }=1,3$ and zero bias%
}

It was shown in Ref.~\onlinecite{yafis1} that, in the Hartree-Fock
approximation, the ground states of the 2DEG in $N=0$ at integer filling
factors satisfy a set of Hund's rules in which the spin polarization is
maximized first, then the layer polarization is maximized to the greatest
extent possible, and finally the orbital polarization is maximized to the
extent allowed by the first two rules. At zero bias, the ordering of the
first four states (with spin up) is given by 

\begin{figure*}[tbph]
\includegraphics[scale=1.0]{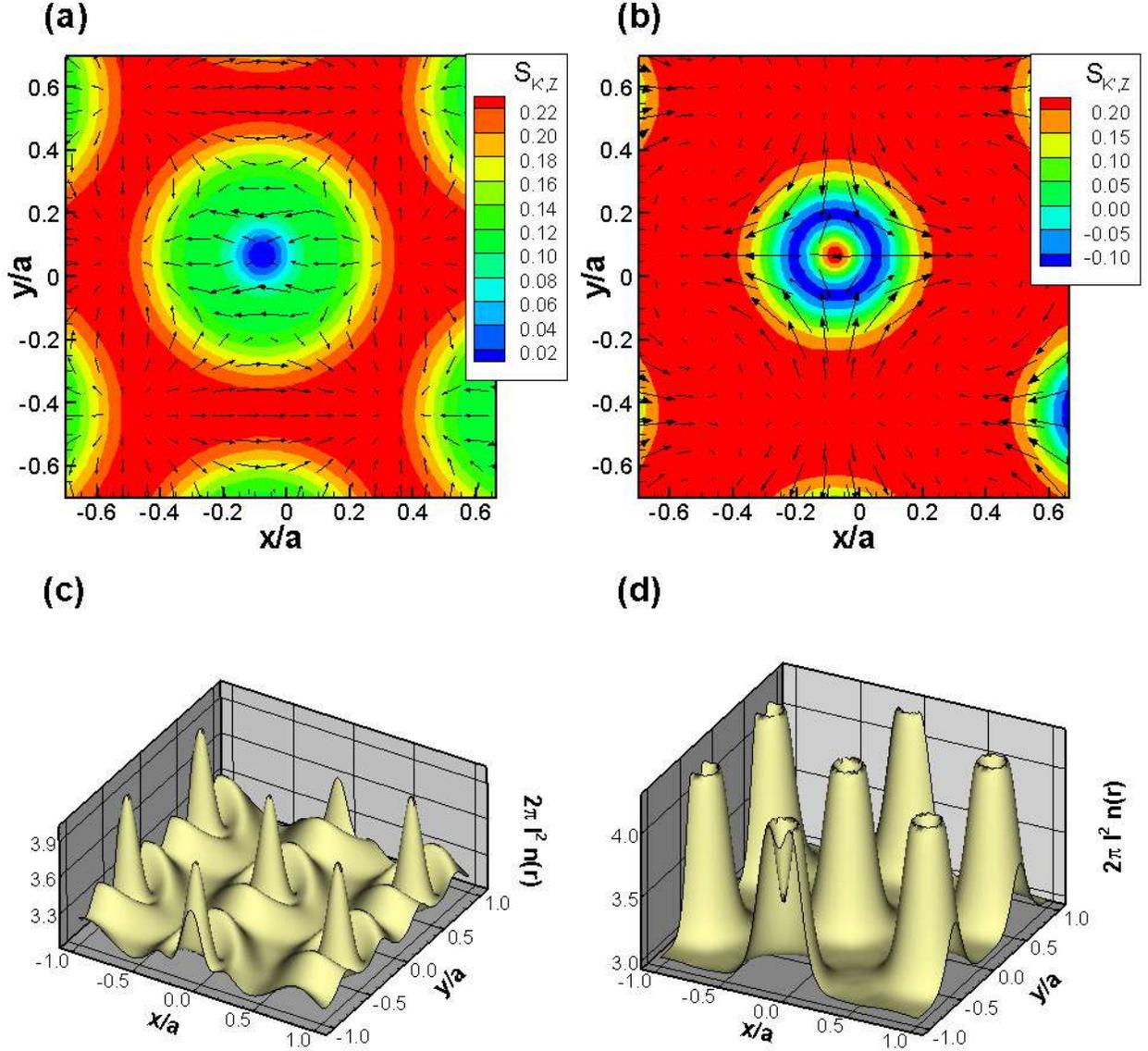}
\caption{(Color online) Orbital Skyrmion
crystal at $\widetilde{\protect\nu }=3.2$ and $\Delta _{B}/\left( e^{2}/%
\protect\kappa \ell \right) =0.$ (a) Orbital pseudospin texture in the RSR.
(b) Orbital pseudospin texture in the GCR. (c) Total density $n\left( \mathbf{r}\right) $ in the RSR. The pseudospin
pattern for $S_{K,Z}$ (not shown) is identical to that of $S_{K^{\prime },Z}$%
. (d) Total density $\widetilde{n}\left( \mathbf{r}\right) $ in the GCR. The lattice
constant is $a$.}
\label{fig7}
\end{figure*}

\begin{eqnarray}
\left\vert S,0\right\rangle &=&\frac{1}{\sqrt{2}}\left\vert K,0\right\rangle
+\frac{1}{\sqrt{2}}\left\vert K^{\prime },0\right\rangle ,  \label{states} \\
\left\vert S,1\right\rangle &=&\frac{1}{\sqrt{2}}\left\vert K,1\right\rangle
+\frac{1}{\sqrt{2}}\left\vert K^{\prime },1\right\rangle ,  \notag \\
\left\vert AS,0\right\rangle &=&\frac{1}{\sqrt{2}}\left\vert
K,0\right\rangle -\frac{1}{\sqrt{2}}\left\vert K^{\prime },0\right\rangle , 
\notag \\
\left\vert AS,1\right\rangle &=&\frac{1}{\sqrt{2}}\left\vert
K,1\right\rangle -\frac{1}{\sqrt{2}}\left\vert K^{\prime },1\right\rangle , 
\notag
\end{eqnarray}%
in this order. (The guiding-center index $X$ is left implicit in these
equations).

At $\widetilde{\nu }=1$, the first level is completely filled while at $%
\widetilde{\nu }=3$, the first three levels are completely filled. At $%
\widetilde{\nu }=1+x$ with $\left\vert x\right\vert \lesssim 0.5$, a finite
density of electrons ($x>0$) or holes ($x<0$), in the $\left\vert
S,1\right\rangle $ or $\left\vert S,0\right\rangle $ state condense into a
crystal phase. At $x=0.2$ and zero bias, we find that the electrons in the $%
\left\vert S,1\right\rangle $ state condense into a triangular crystal with
two orbital Skyrmions per site (i.e. one orbital Skyrmions in each layer).
There is no layer-pseudospin texture (no pseudospin rotation) 
in this case since all electrons are in
a symmetric state of the bilayer but there is an interlayer coherence.
Moreover, an orbital-pseudospin texture is always present.

The crystal state at $\widetilde{\nu }=3.2$ and zero bias is identical to
that at $\widetilde{\nu }=1.2$ with the exception that it is the electrons
in the $\left\vert AS,1\right\rangle $ that now condense into a crystal
phase and orbital Skyrmions are formed by flipping orbital pseudospins from $%
\left\vert AS,0\right\rangle $ to $\left\vert AS,1\right\rangle $. There are
again $2$ orbital Skyrmions per site. The two states $\left\vert
S,0\right\rangle ,\left\vert S,1\right\rangle $ are completely filled and
inert and give a background density of $2/2\pi \ell ^{2}$. The pseudospin
and density patterns in the RSR and GSR for this crystal state with lattice
spacing $a$ are shown in Fig.~\ref{fig7}. (In these figures, $a$ is the Skyrmion lattice
constant and $a>>a_0$).
The pseudospin pattern
for $S_{K,Z} $ (not shown in the figure) is identical to that of $%
S_{K^{\prime },Z.}$ The additional central peak in the RSR density profile
occurs because there are $2$ Skyrmions per site in this crystal. Note that
the pseudospin for the charge $q=2e$ Skyrmion rotates by $4\pi $ in the RSR
but only $2\pi $ in the GCR.

\subsection{Bias $\Delta _{B}>\Delta _{B}^{\left( c\right) }$}

\begin{figure}[tbh]
\includegraphics[scale=1]{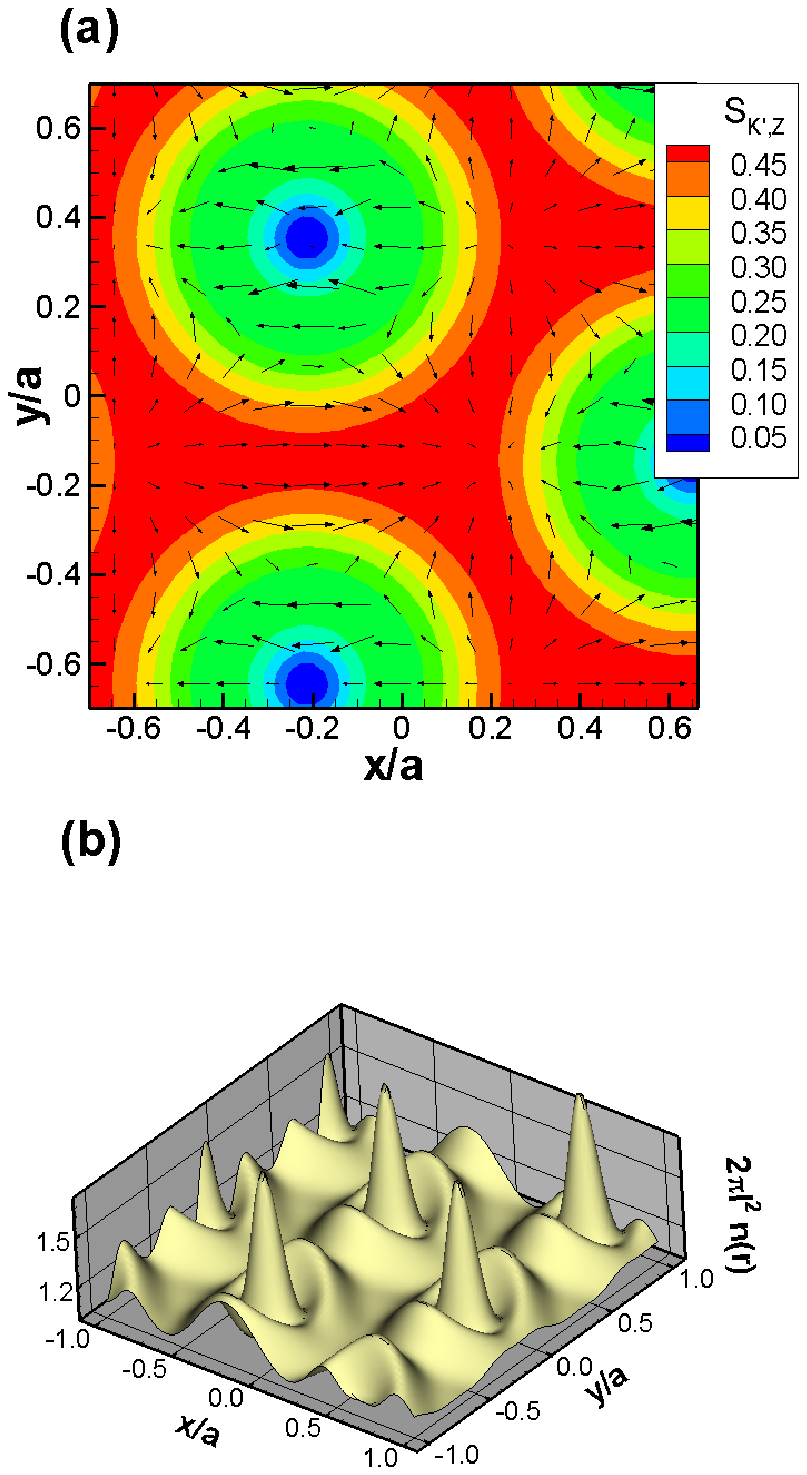}
\caption{(Color online) Orbital Skyrmion crystal at $\widetilde{\protect\nu }%
=1.2$ and $\Delta _{B}/\left( e^{2}/\protect\kappa \ell \right) =0.002$. (a)
Orbital pseudospin texture in the RSR. (b) Total density $n\left( \mathbf{r}%
\right) $ in the RSR. The pseudospin $\mathbf{S}_{K}=0.$}
\label{fig8}
\end{figure}

With finite bias, the charge in layer $K$ is progressively transferred to
layer $K^{\prime }.$ For the crystal states discussed above, that means that
the size of the Skyrmions decreases in layer $K$ and increases in layer $%
K^{\prime }.$ In our mean-field approximation, the charge of the Skyrmions
is not quantized and the crystal states can be seen as exotic charge density waves
with complex pseudospin textures. Above a very small bias of order $\Delta
_{B}^{\left( c\right) }/\left( e^{2}/\kappa \ell \right) \approx 0.0011$ at $%
\widetilde{\nu }=1$, all electrons are pushed into the $\left\vert K^{\prime
},0\right\rangle $ valley and interlayer coherence is lost. The ordering of
the energy levels at $\widetilde{\nu }=1$ is then given by $\left\vert K^{\prime
},0\right\rangle ,\left\vert K^{\prime },1\right\rangle ,\left\vert
K,0\right\rangle ,\left\vert K,1\right\rangle $ and the ground state has all
electrons in $\left\vert K^{\prime },0\right\rangle $. At $\widetilde{\nu }%
=1.2$ and $\Delta _{B}/\left( e^{2}/\kappa \ell \right) =0.002,$ we find a
triangular crystal of orbital Skyrmions in layer $K^{\prime }$ with again
two electrons per site. There are no electrons in valley $K.$ The
guiding-center density and vector fields patterns for this state are given
in Fig.~\ref{fig8}. The total density is identical to that of the crystal
state found at zero bias (see Fig.~\ref{fig7}) but it is now completely in
layer $K^{\prime }$ instead of being equally shared between the two layers.
The orbital pseudospin pattern in $K^{\prime }$ is the sum of the orbital
pseudospin patterns found in each layer at zero bias.

The pseudospin texture and density for a single orbital Skyrmions of charge $%
q=2e$ on each site of the lattice in Fig.~\ref{fig7} can be obtained with
the microscopic expression%
\begin{equation}
\left\vert \mathcal{S}^{\left( 2e\right) }\right\rangle
=\prod\limits_{m=0}^{\infty }\left[ -u_{m}c_{1,m+2}^{\dag
}+v_{m}c_{0,m}^{\dag }\right] c_{1,0}^{\dag }c_{1,-1}^{\dag }\left\vert
0\right\rangle .
\end{equation}%
The density in this state is given by 
\begin{eqnarray}
n_{1}^{\left( 2e\right) }\left( \mathbf{r}\right) &=&\left\vert \phi
_{1,-1}\left( \mathbf{r}\right) \right\vert ^{2}+\left\vert \phi
_{1,0}\left( \mathbf{r}\right) \right\vert ^{2} \\
&&+\sum_{m=0}^{\infty }\left\vert h_{1,m+2}\left( \mathbf{r}\right)
\right\vert ^{2}\left\vert v_{m}\right\vert ^{2},  \notag
\end{eqnarray}%
while the pseudospin texture is still given by Eq.~(\ref{spin1}) and 
\begin{equation}
S_{z}^{\left( 2e\right) }\left( \mathbf{r}\right) =\frac{1}{2}\left[
n_{0}\left( \mathbf{r}\right) -n_{1}^{\left( 2e\right) }\left( \mathbf{r}%
\right) \right] .
\end{equation}

The phase diagram of the liquid state at $\widetilde{\nu }=3$ is richer than
at $\widetilde{\nu }=1$ since a mixed state with both orbital and interlayer
coherences is possible\cite{cote1} due to the fact that the kinetic energy
contribution $-\beta _{0}\Delta _{B}$ in Eq.~(\ref{s_11}) is negative and so
decreases the exchange-enhanced gap $\Delta ^{\ast }=E_{n=0}<E_{n=1}$ with
increasing bias. (This gap $\Delta ^{\ast }$ is positive at zero bias but
changes sign at sufficiently high bias). In the region with orbital
coherence only, i.e. for $\Delta _{B}/\left( e^{2}/\kappa \ell \right)
\gtrsim 0.0022$, the ordering of the levels is given by $\left\vert
K^{\prime },0\right\rangle ,\left\vert K^{\prime },1\right\rangle
,\left\vert K,B\right\rangle ,\left\vert K,AB\right\rangle $ where $B$ and $%
AB$ represent bonding and anti-bonding combinations of the $n=0$ and $n=1$
states defined by

\begin{eqnarray}
\left\vert K,B\right\rangle &=&\sqrt{1-\sigma }\left\vert K,0\right\rangle +%
\sqrt{\sigma }\left\vert K,1\right\rangle ,  \label{b1} \\
\left\vert K,AB\right\rangle &=&-\sqrt{\sigma }\left\vert K,0\right\rangle +%
\sqrt{1-\sigma }\left\vert K,1\right\rangle ,  \label{b2}
\end{eqnarray}%
with%
\begin{equation}
\sigma =\frac{\Delta _{B}}{\Delta _{B}^{(2)}},
\end{equation}%
and 
\begin{equation}
\Delta _{B}^{(2)}=\frac{1}{4\beta }\sqrt{\frac{\pi }{2}}\approx 5\frac{e^{2}%
}{\kappa \ell }
\end{equation}%
(at $B=10$ T) is the bias at which the electrons are completely transferred
to the state $\left\vert K,1\right\rangle $ at $\widetilde{\nu }=3$. The two
levels $\left\vert K^{\prime },0\right\rangle ,\left\vert K^{\prime
},1\right\rangle $ are completely filled. The number of electrons in state $%
\left\vert K,1\right\rangle $ is given by

\begin{equation}
\nu _{K,1}=\frac{\Delta _{B}}{\Delta _{B}^{(2)}}.  \label{m2}
\end{equation}%
At $\widetilde{\nu }=3.2$ and $\Delta _{B}\gtrsim \Delta _{B}^{\left(
c\right)}$ (with $\Delta_{B}^{\left(c\right)}=0.0021$, we get a triangular crystal of orbital Skyrmions with charge $%
q=-2e$ per site with a density and pseudospin patterns in the $n=0,1$ basis
similar to those represented in Fig.~\ref{fig8}. The only difference with
the $\widetilde{\nu }=1.2$ case is that Skyrmions are now made from a filled 
$\left\vert K,B\right\rangle $ level by flipping pseudospins to the $%
\left\vert K,AB\right\rangle $ level. When $\Delta _{B}$ is close to $\Delta
_{B}^{\left( c\right) }$, however, electrons are in majority in level $n=0$
and there is not much difference with the $\widetilde{\nu }=1.2$ case. When
the bias is sufficiently strong for the exchange-enhanced gap to be
negative,the crystal phase is much more complex but the orbital pseudospin
texture persists. We do not discuss this limit further in this paper.

Skyrmion with charge $q=-2e$ are not unknown. It was shown in Ref.~\onlinecite{nazarov}, 
for example, that at small density, there is an
attractive force between two Skyrmions with opposite global phases of their
spin component that goes like $1/R$ where $R$ is the separation between the
two Skyrmions. At large separation $R,$ this force prevails over Coulomb
repulsion. Also, in previous studies of spin and pseudospin Skyrmions in
conventional semiconductor's 2DEG, it was found that lattice with pairs of
Skyrmions occurred for small value of the Zeeman or bias couplings\cite{cotecp3}.

\subsection{Skyrme crystals near $\widetilde{\protect\nu }=2$}

At $\widetilde{\nu }=2,$ the uniform ground state at zero bias has
interlayer coherence in $n=0$ \textit{and} in $n=1$ so that the first two
states in Eqs.~(\ref{states}) are filled. Above a critical bias of the order
of $\Delta _{B}^{(c)}/\left( e^{2}/\kappa \ell \right) \approx 0.003$, all
charge are transferred to the $K^{\prime }$ valley and states $n=0$ and $n=1$
are then fully occupied. Interlayer as well as orbital coherences are lost.

\begin{figure}[tbh]
\includegraphics[scale=1]{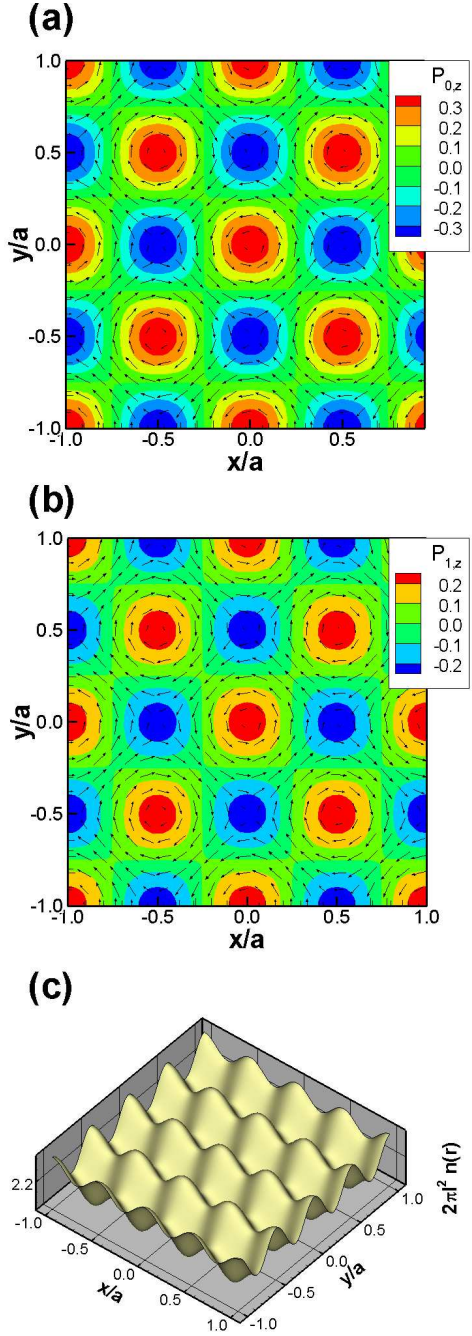}
\caption{(Color online) Interlayer Skyrmion crystal at $\widetilde{\protect\nu }%
=2.2$ and $\Delta _{B}/\left( e^{2}/\protect\kappa \ell \right) =0.$(a)
Layer-pseudospin texture in $n=0$ and (b) Layer-pseudospin texture $n=1$ in
the RSR. (c) Total density $n\left( \mathbf{r}\right) $ in the RSR.}
\label{fig9}
\end{figure}

At $\widetilde{\nu }=2.2$ and zero bias, the electronic phase consists of
two layer-pseudospin meron crystals with each meron carrying charge $q=-e/2.$
There is a meron texture in $\mathbf{P}_{0}$ and in $\mathbf{P}_{1}.$ This
phase is depicted in Fig.~\ref{fig9}. The merons are arranged in a
checkerboard configuration with $8$ merons per unit cell (the total charge
in one unit cell is $4e$). The layer-pseudospins in the central meron are
rotated by a phase $\pi $ with respect to the merons at the corner of the
unit cell. The orbital coherence is more than ten time smaller than the
interlayer coherence and can be neglected.

\begin{figure}[tbph]
\includegraphics[scale=1]{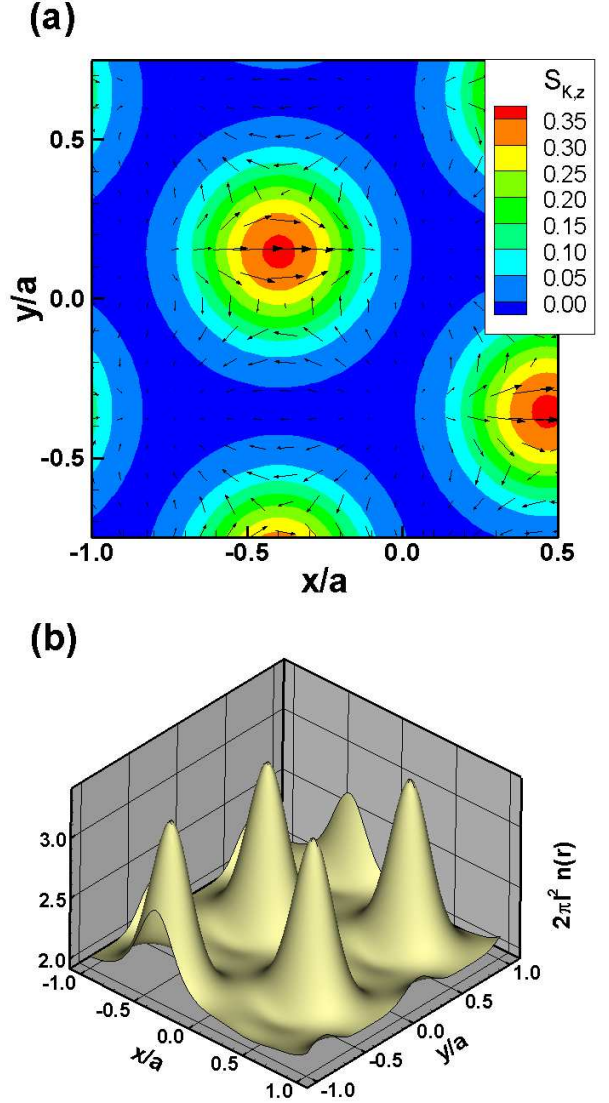}
\caption{(Color online) Orbital crystal at $\widetilde{\protect\nu }=2.2$ and 
$\Delta _{B}/\left( e^{2}/\protect\kappa \ell \right) =0.007.$ (a) Orbital
pseudospin texture in the RSR. (b) Total density $n\left( \mathbf{r}\right) $
in the RSR.}
\label{fig10}
\end{figure}

The charge in the pseudospin merons with pseudospin down (up) at the center
progressively decreases (increases) when the bias is increased. For $%
\widetilde{\nu }=2.2$ and at $\Delta _{B}/\left( e^{2}/\kappa \ell \right)
\approx 0.007$, we find that there is a phase transition to a state where
the two states in valley $K^{\prime }$ are completely filled and the
remaining electrons crystallize in the $K$ valley. In the $n=0,1$ basis,
this gives a triangular crystal with one electron per site and an orbital
pseudospin vortex around each electron. Note that the liquid phase at $%
\widetilde{\nu }=2$ has all electrons in a bonding state $\left\vert
K,B\right\rangle $ of Eq.~(\ref{b1}) (with $\Delta _{B}^{(2)}=(\widetilde{%
\nu }-1)\sqrt{\pi /2}/4\beta $ in this case) and we could have expected a
Wigner crystal phase with electrons in the $\left\vert K,B\right\rangle $
state at each site. It seems however that the system again prefers to form a
pseudospin texture at each site. This crystal state is represented in Fig.~\ref{fig10}.

As in the $\widetilde{\nu }=3.2$ case discussed above, the exchange-enhanced
gap $\Delta ^{\ast }$ between $n=0$ and $n=1$ is of the order of $%
e^{2}/\kappa \ell $ at $\widetilde{\nu }=2.$ The contribution $-\beta
_{0}\Delta _{B}$ (see Eq.~(\ref{s_11})) decreases this gap as $\Delta _{B}$
is increased. This increases the number of flipped orbital pseudospins. At $%
\Delta _{B}^{(2)}=(\widetilde{\nu }-1)\sqrt{\pi /2}/4\beta ,$ there is a
transition to a Wigner crystal with all electrons in $n=1$ at each site and
there is no pseudospin texture anymore.

\begin{figure}[h!]
\includegraphics[scale=1]{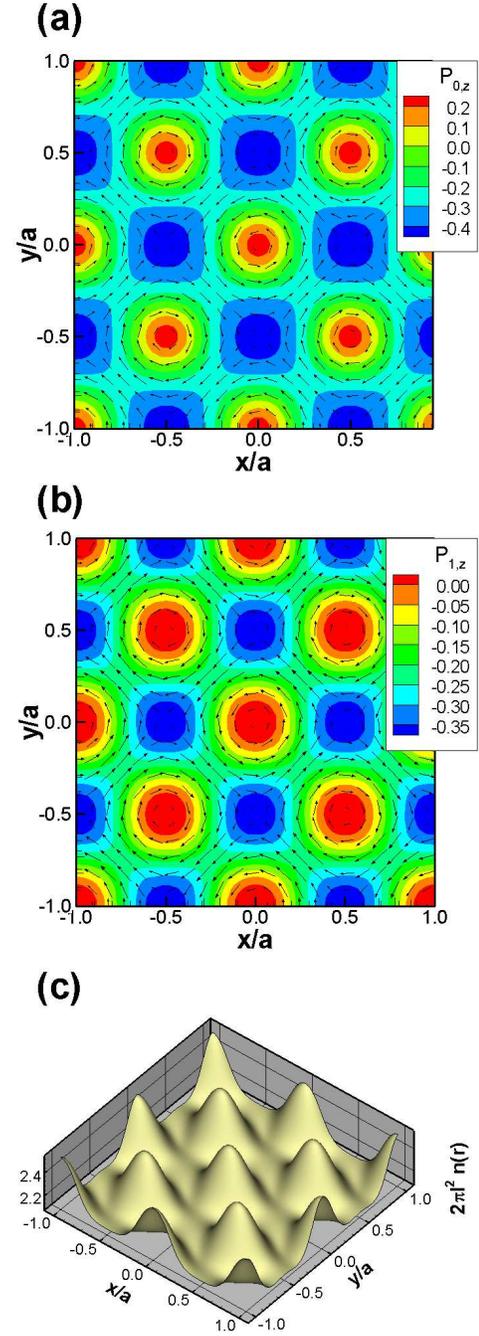}
\caption{(Color online) Interlayer-Skyrmion crystal at $\widetilde{\protect%
\nu}=2.2$ and $\Delta _{B}/\left( e^{2}/\protect\kappa \ell \right) =0.03.$%
(a) Layer-pseudospin texture in $n=0$ and (b) $n=1$ in the RSR. (c) Total
density $n\left( \mathbf{r}\right) $ in the RSR. Note that the majority
state is pseudospin down. This state is not the ground state in our
numerical calculation.}
\label{fig11}
\end{figure}

It is possible to find another interesting solution with our numerical code
at $\Delta _{B}>\Delta _{B}^{(c)}$ that has, however, an higher energy than
that of Fig.~\ref{fig10}. We mention it here because it is closely related
to the work reported in Ref.~\onlinecite{abanin}. This solution is a crystal
of layer-pseudospin Skyrmions in $n=0$ and in $n=1$ with a charge $q=-2e$
Skyrmion at each crystal site. This solution is the natural extension of the
solution at zero bias since the bias transfers the charge of half the merons 
in one layer to the other half in the same layer. This structure is shown
in Fig.~\ref{fig11}. We remark that our convention for the \textit{interlayer%
} pseudospin is that state up corresponds to valley $K.$ Since the charge is
pushed in valley $K^{\prime }$ with positive bias, the majority state is
pseudospin down. In that case, a Skyrmion has spin up at the center and spin
down away from the center.

This type of solution i.e. Skyrmions with a superposition of $n=0$ and $n=1$
interlayer textures have been studied by Abanin et al.\cite{abanin} (the
small contribution $\beta _{0}\Delta _{B}$ to the gap was set to zero in
that paper). These authors concluded that such charge $q=-2e$ Skyrmions
would have lower energy than electron or hole quasiparticles at filling
factor $\widetilde{\nu }=2.0.$ The crystal structure that we get is
consistent with their finding but it is not the ground state. As we mentioned
before, our conclusions for the crystal state do not necessarily apply to the case
of an isolated Skyrmion. 
If we define
a Skyrmion creation operator in states $n=0,1$ as 
\begin{equation}
d_{0}^{\dag }=\prod\limits_{m=0}^{\infty }\left[ -u_{0,m}c_{K,0,m+1}^{\dag
}+v_{0,m}c_{K^{\prime },0,m}^{\dag }\right] c_{K,0}^{\dag }\left\vert
0\right\rangle ,
\end{equation}%
\begin{equation}
d_{1}^{\dag }=\prod\limits_{m=-1}^{\infty }\left[ -u_{1,m}c_{K,1,m+1}^{\dag
}+v_{1,m}c_{K^{\prime },1,m}^{\dag }\right] c_{K,-1}^{\dag }\left\vert
0\right\rangle ,
\end{equation}%
then the $2e$ Skyrmion state can be written as 
\begin{equation}
\left\vert \mathcal{S}^{\left( 2e\right) }\right\rangle =d_{1}^{\dag
}d_{0}^{\dag }\left\vert 0\right\rangle
\end{equation}%
and the angular momentum pairing is such that $m_{K^{\prime }}-m_{K}=1$ for
both Skyrmions.

In a previous publication\cite{cote1}, we derived an effective model for the
orbital pseudospin-wave excitations at $\widetilde{\nu }=3.$ This effective
model had in it a Dzyaloshinskii-Moriya interaction. This type of
interaction favors the formation of spiral or vortex states. We see that
this term is also at work in the crystal states.

\section{TOTAL AND LOCAL DENSITY OF STATES}

Skyrmion lattices with charge $q=-2e$ can be distinguished from Skyrmion
lattices with charge $q=-e$ by their total density of states (TDOS) which is
defined by 
\begin{eqnarray}
g_{T}\left( \omega \right) &=&-\frac{1}{\pi }\sum_{n,a}\int d\mathbf{r}\,%
\mathrm{Im}\left[ G_{n,n}^{\left( R\right) a,a}\left( \mathbf{r},\mathbf{r}%
,\omega \right) \right] \\
&=&-\frac{N_{\phi }}{\pi }\sum_{n,a}\mathrm{Im}\left[ G_{n,n}^{\left( R\right)
a,a}\left( \mathbf{q}=0,\omega \right) \right] ,  \notag
\end{eqnarray}%
where $G_{n,n}^{\left( R\right) a,a}$ (with $a=K,K^{\prime }\,$and $n=0,1$)
is the retarded single-particle Green's function which is related to the
Matsubara Green's function defined in Eq.~(\ref{gmatsubara}) by $%
G_{n,n}^{a,a}\left( \mathbf{q},i\omega _{n}\rightarrow \omega +i\delta
\right) =G_{n,n}^{\left( R\right) a,a}\left( \mathbf{q},\omega \right) .$

\begin{figure}[tbph]
\includegraphics[scale=1]{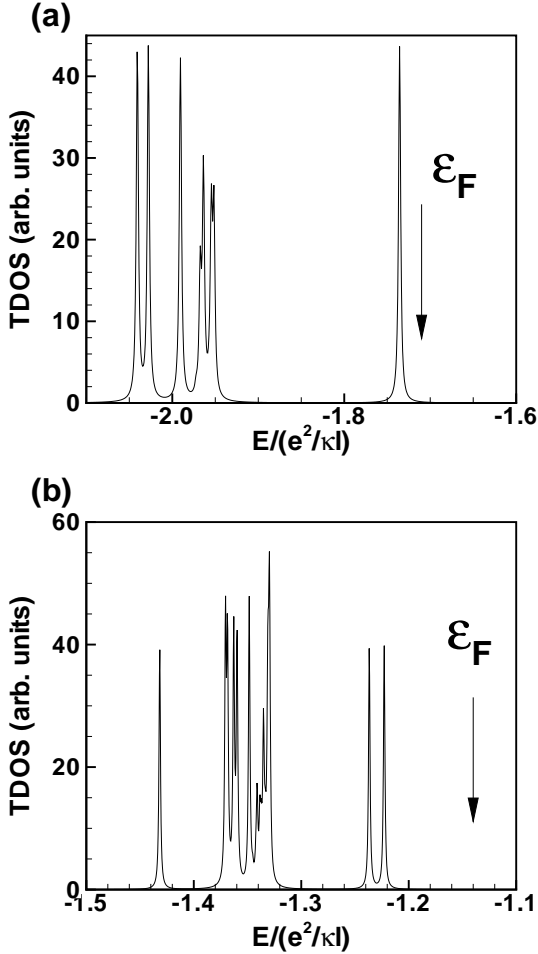}
\caption{Total density of states for orbital Skyrmion crystals at $%
\widetilde{\protect\nu}=1.2$. (a) For $\Delta _{B}/\left( e^{2}/\protect%
\kappa \ell \right) =1.28.$ Skyrmions with $q=-e$. (b) For $\Delta _{B}/\left( e^{2}/\protect\kappa %
\ell \right) =0.002.$ Skyrmions with $q=-2e$. Only the low-energy part of the TDOS is shown in
these figures.}
\label{fig12}
\end{figure}

The number of peaks near the Fermi level in the TDOS is equal to the number
of electrons in a Skyrmion. This is illustrated in Fig.~\ref{fig12} where we
show the low-energy part of $g_{T}\left( \omega \right) $ corresponding to
the crystals of Fig.~\ref{fig6} ($q=-e$ Skyrmion at each site) and Fig.~\ref%
{fig7} ($q=-2e$ Skyrmion at each site). A similar result was also found for
bubble crystals in semiconductor's 2DEG\cite{cotebubble}.

\begin{figure}[tbph]
\includegraphics[scale=1]{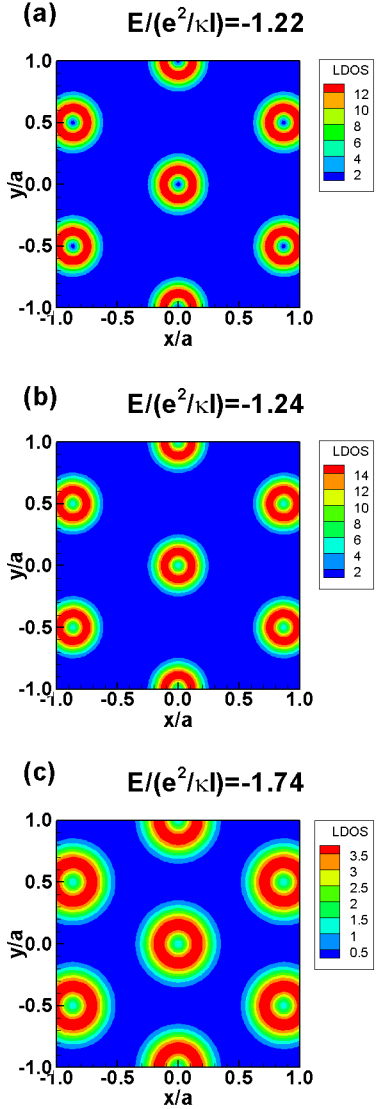}
\caption{(Color online)\ Local density of state (LDOS) in valley $K^{\prime }
$ for the orbital Skyrmion crystals considered in Fig.~\ref{fig12} (a) and (b). (a)
and (b) LDOS for the two higher-energy peaks of the crystal with 2 electrons
per site. (c) LDOS at the higher-energy peak for the crystal with one
electron per site.}
\label{fig13}
\end{figure}

It was shown by Poplavskyy et al.\cite{goerbig} that the density pattern in
the bubble crystal can also be seen by scanning tunneling microscopy. This
measure is related to the \textit{local} density of states (LDOS)\ which is
defined by%
\begin{eqnarray}
g_{L}\left( \mathbf{r},\omega \right) &=&-\frac{1}{\pi }\,\sum_{n,a}\mathrm{Im}%
\left[ G_{n,n}^{\left( R\right) a,a}\left( \mathbf{r},\mathbf{r},\omega
\right) \right] , \\
&=&-\frac{1}{\pi S}\sum_{n,a}\sum_{\mathbf{q}}\mathrm{Im}\left[ \widehat{G}%
_{n,n}^{\left( R\right) a,a}\left( \mathbf{q,}\omega \right) e^{-i\mathbf{q}%
\cdot \mathbf{r}}\right] ,  \notag
\end{eqnarray}%
where 
\begin{eqnarray}
\widehat{G}_{n,n}^{a,a}\left( \mathbf{q,}\omega \right) &\equiv &\int d%
\mathbf{r}e^{i\mathbf{q}\cdot \mathbf{r}}G_{n,n}^{\left( R\right) a,a}\left( 
\mathbf{r},\mathbf{r},\omega \right) , \\
&=&N_{\phi }G_{n,n}^{a,a}\left( -\mathbf{q},\omega \right) K_{n,n}\left( 
\mathbf{q}\right) .  \notag
\end{eqnarray}%
We show in Fig.~\ref{fig13} the LDOS in valley $K^{\prime }$ evaluated at
the energy of the two highest-energy peaks in Fig.~\ref{fig12}(a) and at the
highest-energy peak in Fig.~\ref{fig12}(b). The LDOS\ is almost the same for
both peaks in the case of the Skyrmion crystal with charge $q=-2e.$ The LDOS
for the Skyrmion crystal with charge $q=-e$ looks much the same. Following
Ref.~\onlinecite{goerbig}, we can also sum the LDOS\ evaluated at all the
peaks below the Fermi energy. It is easy to show analytically, using Eq.~(
\ref{g0}), that this summation gives%
\begin{equation}
\int_{-\infty }^{E_{F}}g_{L}\left( \mathbf{r},\omega \right) d\omega
=N_{p}\left( \mathbf{r}\right) ,
\end{equation}%
where $N_{p}\left( \mathbf{r}\right) $ is the density we defined in Sec.~V(b). 
This density is actually quite close to the real space density $%
n\left( \mathbf{r}\right) $ that we plotted in many of the figures of this
paper (it does not contain the term $2\mathrm{Re}\left[ S_{-}\left( \mathbf{r}%
\right) \right] $).

\section{ELECTRIC DIPOLE TEXTURES}

Apart from the minus sign in front of $\left\langle \rho _{a,y}\left( 
\mathbf{G}\right) \right\rangle $ in Eq.~(\ref{dipole}), the vector field
representation for the electric dipoles in the crystal states with orbital
coherence is just like the GCR of the orbital pseudospin field $\mathbf{S}%
_{a,\bot }\left( \mathbf{r}\right) $ where $a=K,K^{\prime }$. We give an
example of the dipole field in Fig.~\ref{fig7} for the charge $q=-2e$
orbital Skyrmion crystal at $\widehat{\nu }=3.2$ and bias $\Delta
_{B}/\left( e^{2}/\kappa \ell \right) =0.$ The rotation of the pseudospins
is $2\pi $ for both charge $q=-e$ and charge $q=-2e$ Skyrmions so that the $%
\mathbf{S}_{a,\bot }$ field pattern does not allow to discriminate between
these two types of crystals.

From Eq.~(\ref{edipolep}), we see that in the presence of an external uniform
electric field $\mathbf{E}=E_{0x}\widehat{\mathbf{x}}+E_{0y}\widehat{\mathbf{%
y}}$ in the plane of the layers, the coupling with the electron gas is given
by%
\begin{eqnarray}
H_{ext} &=&\sqrt{2}\ell eN_{\varphi }\int d\mathbf{r}\left( E_{0,x}\overline{%
\rho }_{j,x}\left( \mathbf{r}\right) -E_{0,y}\overline{\rho }_{j,y}\left( 
\mathbf{r}\right) \right) \\
&=&-\sqrt{2}\ell eN_{\varphi }\left( E_{0,x}\rho _{j,x}\left( \mathbf{G=0}%
\right) -E_{0,y}\rho _{j,y}\left( \mathbf{G=0}\right) \right) .  \notag
\end{eqnarray}%
In the liquid phase where the orientation of the orbital pseudospins in the $%
x-y$ plane is arbitrary, this term allows us to rotate the orbital
pseudospins in that plane. For a Skyrmion crystal, the effect may be more
complex. The parallel electric field forces the orbital pseudospin in the $%
x-y$ plane and so should increase orbital coherence. It should also change
the form of the orbital pseudospin texture of the Skyrmions, orienting them
more towards the $x$ axis. 

\section{CONCLUSION}

We have presented in this paper a study of some crystal phases with valley
and/or orbital pseudospin textures that can occur in bilayer graphene away
from integer filling factors in Landau level $N=0.$ Our calculations are,
strictly speaking, valid within the two-band tight-binding model introduced
in Sec.~II and within the Hartree-Fock approximation.

In our numerical calculatons, we have neglected the terms $\beta _{4}$ and $%
\Delta $ in Eq.~(\ref{s_3}). These terms change the gap between the two
orbital states $n=0$ and $n=1$ to $\zeta _{1}-\beta _{0}\Delta _{B}$ in the $%
K$ valley and to $\zeta _{1}+\beta _{0}\Delta _{B}$ in the $K^{\prime }\ $%
valley (see Fig.~\ref{fig2}). As we mentioned in Sec.~II, the value of $\zeta
_{1}$ is not known precisely. If we take the values for $\beta _{0},\beta
_{4}$ and $\Delta $ cited in Sec.~II, we find $\zeta _{1}/\left(
e^{2}/\kappa \ell \right) =0.113$ at $B=10$ T so $\zeta _{1}$ is probably
not small. Since the critical bias needed to push the charge in one layer is
such that $\beta _{0}\Delta _{B}^{(c)}/(e^2/\kappa\ell)\approx 0.177\times 10^{-3},$ we see
that these additional terms have the possibility to change the phase diagram
in an important way especially the phases at small or zero bias.
Furthermore, the orbital coherence depends on the gap between the two
orbital state. With the value of $\zeta _{1}$ cited above, the bias $\Delta
_{B}$ needed to place the orbital $n=1$ below $n=0$ in the valley $K$ is $%
\Delta _{B}\approx 1.27$ i.e. a large value. Fortunately, our numerical
calculations show that the orbital Skyrmions crystal (which is the most
important state we discussed in this paper) does survive in the phase diagram
even with a finite $\zeta _{1}$. At filling factor $\widetilde{\nu }=1.2$,
the additional gap suppress the $q=-2e$ orbital Skyrmion crystal in favor of
a $q=-e$ orbital Skyrmion crystal. This is consistent with our mention in
Sec.~VI (b) that Skyrmion with $q=-2e$ are found at small gap. At filling
factor $\widetilde{\nu }=3.2,$ the gap $\zeta _{1}-\beta _{0}\Delta _{B}$
can be made small (or even negative) and orbital Skyrmion crystals of both
types are found.

Skyrmion crystals have both phonon and spin (or pseudospin)\ wave modes. In
Ref.~\onlinecite{fluctuations}, it was shown that the classical (or quantum
mean-field) energy of the Skyrmion is independent of the angle $\varphi $
which defines the global $XY$ orientation of the spin components. This extra 
$U\left( 1\right) $ degree of freedom for a single Skyrmion leads to a
broken symmetry in the crystal ground state and hence to a spin wave mode
which remains gapless in the presence of a Zeeman field. We expect a similar
gapless mode for a crystal of orbital Skyrmions. Fluctuations due to phonons
and to these gapless pseudospin modes will have to be considered at finite
temperature in order to evaluate the stability of the crystal structures
discussed in the present paper.

\begin{acknowledgments}
R. C\^{o}t\'{e} was supported by a grant from the Natural Sciences and
Engineering Research Council of Canada (NSERC). Y. Barlas was supported by a
grant from the State of Florida. A.-H. MacDonald was supported by the NSF\
under grant DMR-0606489 and by the Welch Foundation Grant F1473. Computer
time was provided by the R\'{e}seau Qu\'{e}b\'{e}cois de Calcul Haute
Performance (RQCHP).
\end{acknowledgments}

\end{document}